\documentclass[prb,twocolumn,superscriptaddress,showpacs]{revtex4}

\usepackage{graphicx}
\usepackage{epsfig}
\usepackage{epstopdf}	
\usepackage{amssymb,amsmath,amsfonts,hyperref,wasysym}
\usepackage{placeins} 
\usepackage{verbatim} 
\usepackage{color}
\usepackage{longtable}

\begin{document}

\newcommand{\SAVE}[1] {{}}  

\title{Transitions to valence-bond solid order in a honeycomb lattice antiferromagnet}

\author{Sumiran Pujari}
\affiliation{Laboratoire de Physique Th\'eorique, IRSAMC, Universit\'e de Toulouse, CNRS, 31062 Toulouse, France}
\affiliation{Dept. of Physics, University of Kentucky, Lexington, KY 50406, USA.}
\author{Fabien Alet}
\affiliation{Laboratoire de Physique Th\'eorique, IRSAMC, Universit\'e de Toulouse, CNRS, 31062 Toulouse, France}
\author{Kedar Damle}
\affiliation{Department of Theoretical Physics, Tata Institute of Fundamental Research, Mumbai 400 005, India}

\begin{abstract}
We use Quantum Monte-Carlo methods to study the ground state phase diagram of a $S=1/2$ honeycomb lattice magnet
in which a nearest-neighbor antiferromagnetic exchange $J$ (favoring
N\'eel order) competes with two different multi-spin interaction terms: a six-spin interaction $Q_3$ that favors columnar valence-bond solid (VBS) order, and a four-spin interaction $Q_2$ that favors staggered VBS order. For $Q_3 \sim Q_2 \gg J$,
we establish that the competition between the two different VBS orders stabilizes
N\'eel order in a large swathe of the phase diagram even when $J$ is the smallest
energy-scale in the Hamiltonian. When $Q_3 \gg (Q_2,J)$ ($Q_2 \gg (Q_3,J)$), this
model exhibits at zero temperature phase transition from the N\'eel state to
a columnar (staggered) VBS state.
We establish that the N\'eel-columnar VBS transition is continuous for all values of $Q_2$, and that critical
properties along the entire phase boundary are
well-characterized by critical exponents and amplitudes of the non-compact
CP$^1$ (NCCP$^1$) theory of deconfined criticality, similar to what is observed on a square lattice. However, a surprising three-fold anisotropy of the phase of the VBS order parameter at criticality, whose presence was recently noted at the $Q_2=0$ deconfined critical point, is seen to persist all along
this phase boundary. We use a classical analogy to explore this by studying the critical point of a three-dimensional $XY$ model with a four-fold anisotropy field which is known to be weakly irrelevant at the three-dimensional $XY$ critical point. In this case, we again find that the critical anisotropy appears to saturate to a nonzero value over the range of sizes accessible to our simulations.

\end{abstract}
\maketitle


\section{Introduction}
\label{sec:intro}
Ground states of quantum magnets with $S=1/2$ moments on a two dimensional (2d) bipartite lattice (such as square or honeycomb lattices)
generally exhibit long-range spin correlations at the N\'eel wavevector ${\mathbf Q}$\cite{Auerbach_book}. This $T=0$ antiferromagnetic order, encoded in a nonzero value of the N\'eel order parameter vector $\vec{n}$,  can be destroyed by frustrating further-neighbour\cite{J1J2modelonsquarelattice1,J1J2modelonsquarelattice2,Albuquerque,Ganesh,Zhu,Gong,Gong2} or ring-exchange interactions, as well
as by certain more tractable multi-spin couplings designed\cite{DesignerHamiltonian_review} to partially mimic the effect of such frustrating interactions. In many examples, the resulting phase has no magnetic order and instead exhibits spatial ordering of the bond-energy. In such
a bond-ordered valence-bond solid (VBS) state\cite{Sachdev_Vojta_review}, 
the singlet projector $P_{\langle i j\rangle} = -\vec{S}_i \cdot \vec{S}_j +1/4$ of two
nearest-neighbor spins $\langle i j \rangle$ has an expectation
value that exhibits spatial structure at the VBS ordering wave-vector(s) ${\bf K}$, resulting
in a non-zero value for the complex VBS order-parameter $\psi$.

\begin{figure}
\includegraphics[width=0.85\hsize]{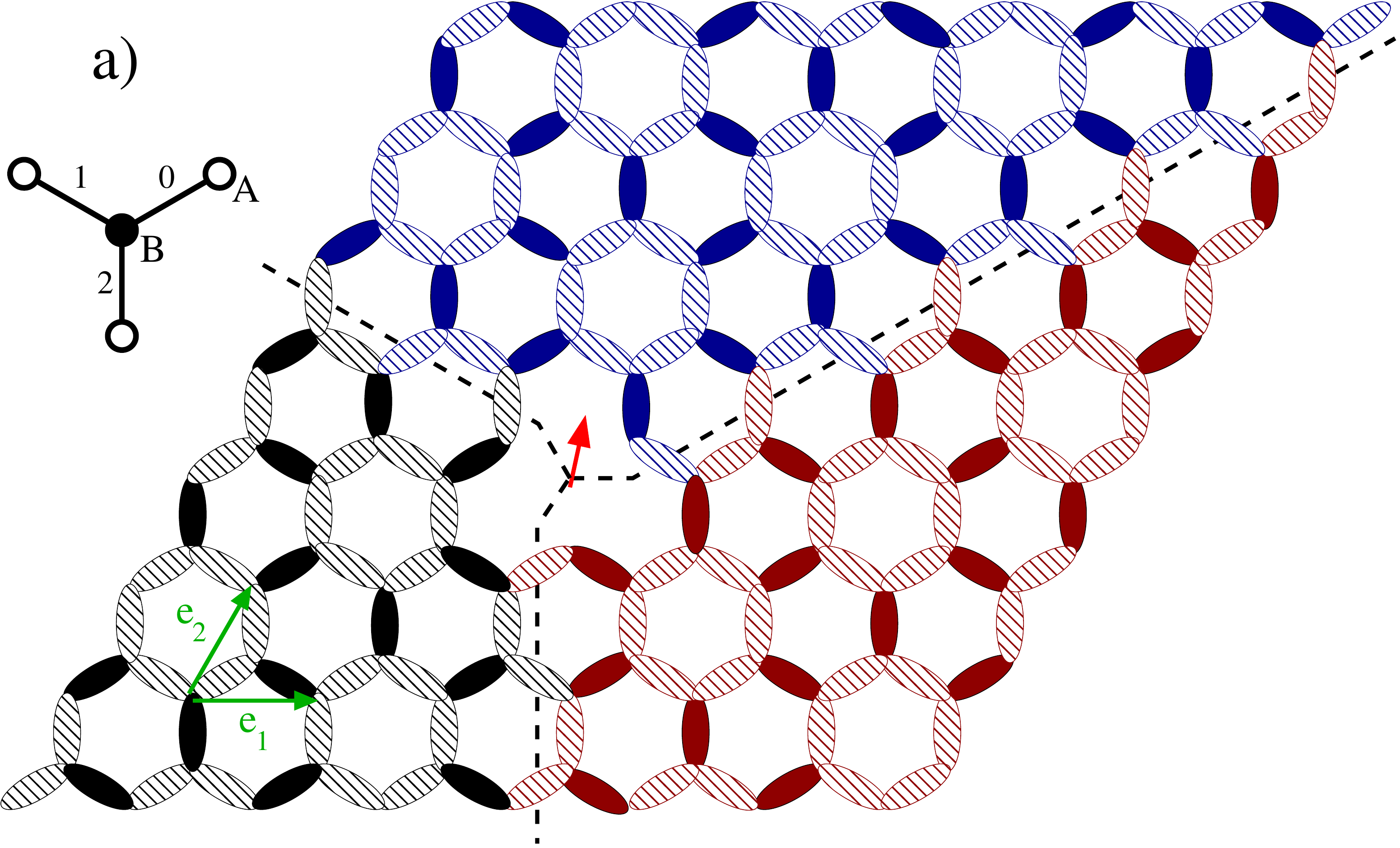}
\includegraphics[width=0.85\hsize]{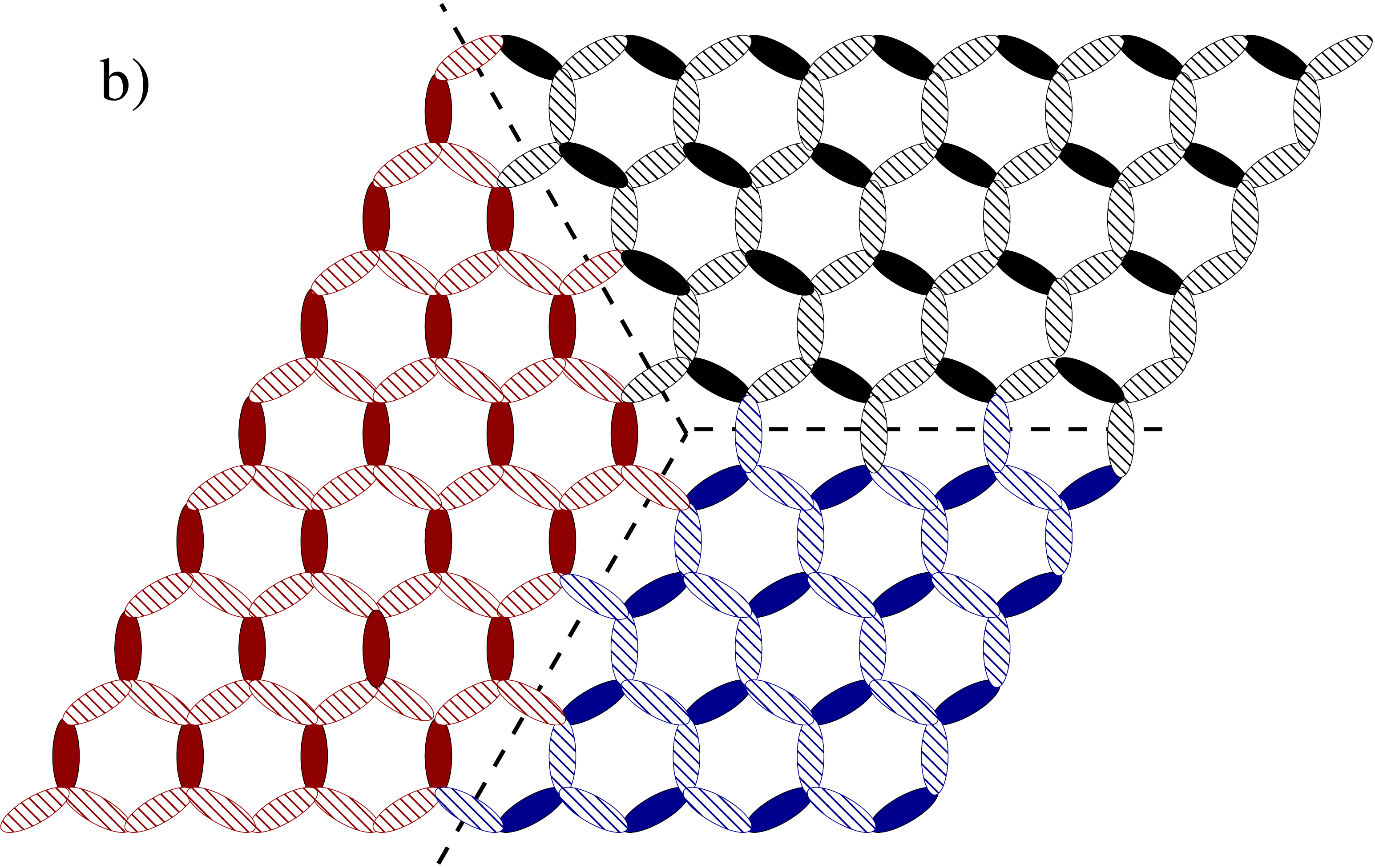}
\includegraphics[width=0.85\hsize]{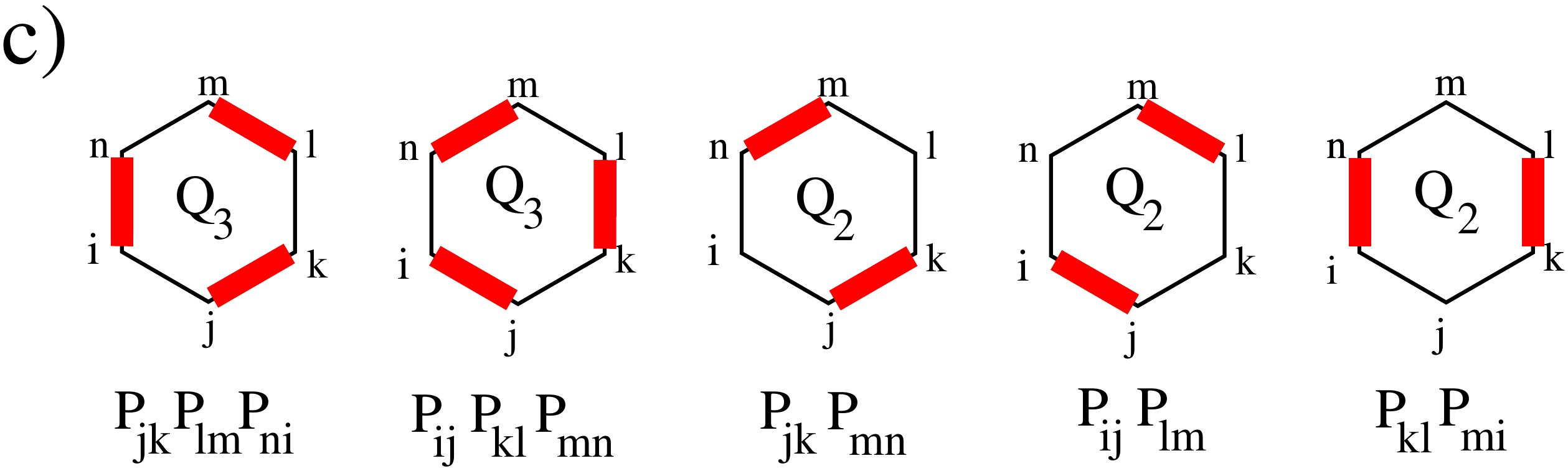}
\caption{(Color online) (a) Columnar VBS order on the honeycomb lattice: dark
links represent higher values of  $\langle P_l \rangle$ (the singlet projection operator on this link) than light links. If dark links are instead reinterpreted as representing
lower values of $P_l$, one obtains a representation of plaquette VBS order
at the same wave-vector. (b) Staggered VBS order on the honeycomb lattice, where again  dark links represent higher values of $\langle P_l \rangle$. In both figures, we have created different ordered domains (represented by different colors and separated by dashed lines) by introducing a defect. As already discussed~\cite{Banerjee_Damle_Paramekanti_2010}, the defect has a spinfull core (a free spin $1/2$ sits at the domain wall intersection) for columnar/plaquette VBS whereas the core is spinless for the staggered VBS.  Also shown are our conventions
for labeling unit cells $\vec{r}$, bonds $\mu$ belonging to unit cells $\vec{r}$, and
$A$ and $B$ sublattice sites in unit cell $\vec{r}$. (c) Schematic representations of the $4-$ and $6-$ spins interactions terms $Q_2$ and $Q_3$.
}
\label{fig:latticeandorders}
\end{figure}

A standard Landau approach (based on a coarse-grained free-energy density\cite{Landau_book} expressed in terms of powers of $\vec{n}$ and $\psi$ and their space-time gradients) would predict that
this phase transformation generically proceeds either via a direct first-order transition, or
via two continuous transitions separated
by an intermediate phase which has both orders or no order. Since the latter possibilities
are more exotic, the simplest generic possibility within Landau theory is thus a direct first-order transition.
Such first-order behavior is indeed observed in square\cite{Sen_Sandvik} and honeycomb lattice\cite{Banerjee_Damle_Paramekanti_2010} spin models where a multi-spin interaction drives the system to a staggered VBS state (Fig.~\ref{fig:latticeandorders} b). 

The theory of deconfined quantum critical points\cite{Senthil_etal_PRB, Senthil_etal_Science,Levin_Senthil} proposed by Senthil {\em et. al.}
argues that such Landau-theory considerations are misleading when
the transition is towards a state with columnar VBS order (Fig.~\ref{fig:latticeandorders}a) on the square
or honeycomb lattice. Indeed, their arguments\cite{Senthil_etal_PRB, Senthil_etal_Science,Levin_Senthil} strongly suggest
that such transitions can be generically (without fine-tuning any parameter) second order in nature. In this alternate approach, one writes the partition function as an imaginary-time ($\tau$) path-integral over space-time configurations $\vec{n}(\vec{r},\tau)$,
 and notes that the spatial configuration
$\vec{n}(\vec{r})$ on a given time-slice admits topological skyrmion textures in spatial
dimension $d=2$. 
The corresponding total skyrmion number is conserved during the imaginary-time evolution as long as the space-time configuration of $\vec{n}$ remains non-singular. Conversely, when the skyrmion number-changing operator $\Psi_{\vec{R}}$ acts at imaginary time
$\tau$ on plaquette $\vec{R}$, it creates a hedgehog defect centered at $\vec{R},\tau$. In this path-integral representation, this hedgehog defect
carries a Berry-phase $2\pi p(\vec{R})/q$ where $p(\vec{R})=0,1\dots q-1$ depends on
the sublattice to which $\vec{R}$ belongs and $q=3$ ($q=4$) for the honeycomb
(square) lattice case\cite{Haldane,Read_Sachdev_PRL89,Read_Sachdev_PRB90,Dadda}. 

Remarkably, this phase factor ensures that the transformation properties of $\Psi$ under lattice symmetries are identical to those of the complex VBS order parameter $\psi$ for columnar order on both honeycomb and square lattices\cite{Read_Sachdev_PRL89,Read_Sachdev_PRB90,Senthil_etal_PRB,Senthil_etal_Science}. The two
operators can thus be identified with each other insofar as their long-distance correlations
are concerned (here and henceforth, we refer to $\psi$ as the ``columnar'' order parameter,
although $\psi$ is also non-zero if the system has plaquette VBS order as shown
in Fig.~\ref{fig:latticeandorders}a for the honeycomb lattice case). The destruction of N\'eel order in the ground state can be described as a proliferation of such hedgehog defects, providing a natural mechanism
for a direct transition between N\'eel and columnar VBS orders\cite{Senthil_etal_PRB, Senthil_etal_Science,Levin_Senthil}. This theoretical description only involves $q$-fold ($q=3$ on the honeycomb lattice and $q=4$ on the square lattice) hedgehogs (corresponding
to $\Psi^q$ and its Hermitian conjugate), as
defects with smaller hedgehog-number carry rapidly oscillating Berry-phases, causing
the corresponding terms in the action to scale to zero upon coarse-graining. Such restrictions on hedgehog charges in space-time configurations of $\vec{n}$ are best analyzed\cite{Motrunich_Vishwanath} in the CP$^1$ representation $\vec{n} = z^{*}_{\alpha}\vec{\sigma}_{\alpha \beta} z_{\beta}$, where $z_\alpha$ is a two-component complex field and $\vec{\sigma}$ the vector of Pauli matrices. In the CP$^1$ representation, hedgehogs correspond to 
monopoles in the compact $U(1)$ gauge-field to which the $z_{\alpha}$ are minimally coupled\cite{Lau_Dasgupta, Kamal_Murthy, Motrunich_Vishwanath}. Thus, if the corresponding {\em non-compact} CP$^1$ theory (NCCP$^1$) has a second-order transition, and if $3$-fold ($4$-fold) monopoles are irrelevant perturbations at the corresponding monopole-free fixed point, one expects that
the N\'eel-columnar VBS
transition on the honeycomb (square) lattice to be generically continuous, with critical properties in the NCCP$^1$ universality class\cite{Senthil_etal_PRB, Senthil_etal_Science,Levin_Senthil}. Conversely, if $3$-fold ($4$-fold) monopoles are relevant at the putative NCCP$^1$ critical point, the simplest
scenario is that this leads to runaway flows which signal weakly-first order behavior
for the N\'eel-columnar VBS transition on the honeycomb (square) lattice\cite{Senthil_etal_PRB, Senthil_etal_Science,Levin_Senthil}.

To understand the scaling behavior of $q$-fold monopole creation operators
in the vicinity of the non-compact CP$^1$ critical point, it is instructive to consider a
more general NCCP$^{N-1}$ theory which has $N$-component fields $z_{\alpha}$ and study the limiting behavior of $q$-fold monopole perturbations in the $N=1$ and $N=\infty$ limits. For instance, four-fold monopoles are known to be irrelevant both at $N=1$\cite{Read_Sachdev_PRB90,Senthil_etal_PRB,Sachdev_Jalabert,Oshikawa,Lou_Sandvik_Balents} and $N=\infty$\cite{Read_Sachdev_PRB90,Senthil_etal_PRB,Sachdev_Jalabert}, making
it very likely that they are also irrelevant in the physical $N=2$ case\cite{Senthil_etal_PRB, Senthil_etal_Science,Levin_Senthil}. Thus, the N\'eel-columnar
VBS transition on the square lattice is expected to be generically second-order, with
critical properties described by the NCCP$^1$ theory\cite{Senthil_etal_PRB, Senthil_etal_Science,Levin_Senthil}. 

The behavior of three-fold monopoles at the noncompact CP$^1$ critical point is harder to 
understand from such a study of limiting cases. This is because the physical $N=2$ case lies between the $N=1$
case where three-fold monopoles are {\em relevant}\cite{Read_Sachdev_PRB90,Senthil_etal_PRB,Sachdev_Jalabert,Oshikawa} and lead to
a {\em weakly-first order transition}\cite{Janke}, and the $N=\infty$\cite{Read_Sachdev_PRB90,Senthil_etal_PRB,Sachdev_Jalabert} limit where they are irrelevant. These contrasting behaviors in the two limits makes it difficult to argue one way
or the other concerning the behavior of three-fold monopole perturbations at the $N=2$ critical point\cite{Senthil_etal_PRB, Senthil_etal_Science,Levin_Senthil}. 
A nice summary of the expected behavior of the NCCP$^{N-1}$ theory with $q$-fold monopoles (including results of numerical simulations) can be found in Ref.~\onlinecite{Block_Melko_Kaul}. 

This theory of deconfined criticality has motivated several numerical studies\cite{Sandvik_PRL2007,Melko_Kaul_PRL2008,Jiang_etal_JStatmech2008,Lou_etal_PRB09,Beach_2009,Sandvik_PRL2010,Banerjee_etal_2010,Kaul_2011,Banerjee_etal_2011,Kaul_Sandvik_PRL2012,Kaul_2012,Sandvik_2012,Jin_Sandvik_2013,Block_Melko_Kaul,Harada_Suzuki_etal,Chen,Pujari_Damle_Alet,Kaul_2014} of model quantum Hamiltonians designed~\cite{DesignerHamiltonian_review} to host a N\'eel-VBS columnar transition. In parallel work, other studies have tried to access the physics
of deconfined criticality in three dimensional classical models~\cite{Chen,Motrunich_Vishwanath2,Kuklov,Sreejith_Powell,Charrier_Alet,Powell_Chalker_2009,Chen_2009,Powell_Chalker_2008,Charrier_Alet_Pujol,Misguich_2008,Alet_2006}. 
On the square lattice (with $q=4$), QMC simulations\cite{Sandvik_PRL2007,Melko_Kaul_PRL2008,Jiang_etal_JStatmech2008,Lou_etal_PRB09,Sandvik_PRL2010,Banerjee_etal_2010,Kaul_2011,Banerjee_etal_2011,Kaul_Sandvik_PRL2012,Sandvik_2012,Chen,Block_Melko_Kaul,Harada_Suzuki_etal,Kaul_2014} find no {\it direct} signature of first-order behavior even
at the largest sizes studied. This is true both for SU(2) symmetric models, as well
as spin models with enhanced SU(N) symmetry, which are expected to
exhibit a transition in the NCCP$^{N-1}$ universality class. Further, critical properties fit reasonably well
to standard scaling predictions for second-order transitions\cite{Sandvik_PRL2007,Sandvik_PRL2010,Banerjee_etal_2010,Kaul_2011,Banerjee_etal_2011,Lou_etal_PRB09,Kaul_Sandvik_PRL2012,Melko_Kaul_PRL2008,Block_Melko_Kaul,Harada_Suzuki_etal}. The corresponding
values of $\eta_N$ and $\eta_D$, the anomalous exponents governing
power-law decays of the N\'eel order parameter $\vec{n}$ and the VBS
order parameter $\psi$, are relatively large~\cite{Sandvik_2012,Block_Melko_Kaul,Harada_Suzuki_etal}, as expected from the theory of
deconfined criticality. Additionally, the numerically estimated critical exponents for large values of $N$ (using lattice spin models with SU($N$) symmetry) approach the limiting values obtained in a large-$N$ expansion of the NCCP$^{N-1}$ theory~\cite{Kaul_Melko_2008,Kaul_Sandvik_PRL2012,Block_Melko_Kaul}. Further, different ``designer Hamiltonians'' with
 different multi-spin couplings~\cite{Sandvik_PRL2007,Sandvik_PRL2010,Kaul_Sandvik_PRL2012} yield the same
estimates for exponents and critical amplitudes.
At or close to this critical point, histograms of the phase of
$\psi$ exhibit near-perfect U(1) symmetry~\cite{Sandvik_PRL2007,Lou_etal_PRB09,Sandvik_2012}, consistent with the idea
that the irrelevance of the $4$-fold monopole insertion operator
$\Psi^4$ implies, via the identification $\Psi \sim \psi$,  the irrelevance of the $4$-fold anisotropy in the phase of the VBS order parameter $\psi$. However, in the SU(2) case,
slow (perhaps
logarithmic) drifts with increasing linear size $L$ are clearly visible\cite{Jiang_etal_JStatmech2008,Sandvik_PRL2010,Banerjee_etal_2010,Kaul_2011,Banerjee_etal_2011,Chen} in certain dimensionless quantities
which are expected to be scale-invariant at a conventional second-order critical point
in three space-time dimensions --- examples include the spin stiffness and vacancy-induced spin textures. Since histograms of phase of $\psi$ exhibit U(1) symmetry characteristic of the non-compact
theory, it seems plausible that these drifts are intrinsic properties of
the non-compact critical point. This interpretation is supported by
the fact that Monte-Carlo simulations of a lattice-regularized
NCCP$^1$ theory\cite{Chen,Motrunich_Vishwanath2,Kuklov} also see some drifts that mar otherwise
convincing scaling behavior (it is also possible
to find different lattice-regularizations that lead to a first-order behavior\cite{Chen,Motrunich_Vishwanath2,Kuklov}). However, at the present juncture, there is no detailed understanding of
these drifts that goes beyond this reasonable guess (see however the recent analytical arguments of Ref.~\onlinecite{Nogueira_Sudbo,Bartosch}).  Finally, we caution that some authors~\cite{Jiang_etal_JStatmech2008,Chen} have also interpreted these drifts as either hints of a very weak first order transition, or as the signature of a flow towards a new universality class different from NCCP$^1$.

What about the honeycomb lattice case ($q=3$)? Recent numerical studies of tractable
model Hamiltonians provide a fairly consistent picture of a direct second-order transition between the N\'eel and the columnar VBS states\cite{Pujari_Damle_Alet,Block_Melko_Kaul,Harada_Suzuki_etal}, with numerical
estimates of the anomalous exponents $\eta_N$ and $\eta_D$, correlation length exponent $\nu$, and universal scaling functions~\cite{Harada_Suzuki_etal} all consistent, within errors, with the best estimates for the square-lattice transition. Further,
slow drifts in spin stiffness analogous to the square lattice case, have also been observed at the putative critical point~\cite{Pujari_Damle_Alet}. All this strongly suggests that the honeycomb lattice transition is also described
by the NCCP$^1$ theory of deconfined criticality. 

However, our recent work has also identified an important new feature of the honeycomb
lattice transition~\cite{Pujari_Damle_Alet}:  if the honeycomb lattice
transition is indeed described by the NCCP$^1$ theory, $3$-fold monopoles
must be irrelevant at the NCCP$^1$ critical point. Since $\Psi \sim \psi$, this
would imply that three-fold anisotropy in the phase of the VBS order parameter $\psi$ is irrelevant at criticality. However, it was found~\cite{Pujari_Damle_Alet} that dimensionless measures of this three-fold anisotropy at criticality appear to saturate to a non-zero value as a function of increasing size (at least for the sizes at which numerical calculations were feasible, which are comparable with those used in square lattice studies). The simplest explanation is that tripled monopoles are irrelevant
with a very small scaling dimension, meaning that the dimensionless critical three-fold anisotropy should flow to zero very slowly. If one only has access to data over a limited range of sizes, it can appear to saturate at a non-zero value.

The present study aims at clarifying this issue of anisotropy, as well as adding some further numerical evidence for the less documented case of deconfined criticality on the honeycomb lattice, relevant for frustrated honeycomb lattice spin models~\cite{Albuquerque,Ganesh,Zhu,Gong}. In this context, we note that a recent study~\cite{Lee_Sachdev} suggests an
interesting experimental realization of deconfined criticality in bilayer graphene in
magnetic and electric fields, further adding to our motivation for studying the N\'eel-columnar VBS transition on the honeycomb lattice. 

We focus here on a numerically tractable model
in which the nearest-neighbor antiferromagnetic exchange $J$ competes with two different multi-spin interaction terms, a six-spin interaction $Q_3$ that favors a columnar VBS state (when $Q_3 \gg Q_2,J$), and
a four-spin interaction $Q_2$ that favors a staggered VBS (when $Q_2 \gg Q_3,J$).
The deconfined quantum critical point for the model at $Q_2=0$ has been studied in our previous work~\cite{Pujari_Damle_Alet}, as well as in Ref.~\onlinecite{Harada_Suzuki_etal}. The motivation for perturbing this model with the $Q_2$ term was three-fold: {\it (i) } this new energy scale (when not too large) will introduce a critical line of $Q_{3c}(Q_2)$ for the N\'eel-columnar VBS
phase boundary. Universality of critical exponents and amplitudes can be tested along this critical line~\footnote{Ref.~\onlinecite{Kaul_2014} recently found a similar critical N\'eel-VBS line on the square lattice, but did not test for universality.}; {\it (ii)} if $Q_2$ tunes the ``bare'' value of the
three-fold anisotropy of the columnar VBS order parameter $\psi$, one could test the behavior of the critical three-fold anisotropy along the phase boundary line $Q_{3c}(Q_2)$; {\it (iii)} the competition between the staggered and columnar VBS orders in the regime
$Q_2 \sim Q_3 \gg J$ may reveal exotic physics: the transition from staggered VBS order (with maximal winding in the valence-bond pattern) to columnar VBS order (with zero-winding) may proceed through an intervening quantum spin-liquid (where no
winding sector is favored). 

Before proceeding further, it is useful to summarize the key findings of the present work:
{\it (i)} we establish that the transition from N\'eel to columnar
VBS order is continuous for all values of $Q_2$, and that critical
properties along the entire N\'eel-columnar VBS phase boundary $Q_{3c}(Q_2)$ are
well-characterized by critical exponents and amplitudes of the NCCP$^1$ theory of deconfined criticality; {\it (ii) } the three-fold anisotropy of the phase of the VBS order parameter persists all along
this phase boundary, with slight but perceptible {\it upward} drift in
its value as $Q_2$ is increased. To explore the possibility that this may reflect the fact that tripled-monopoles are irrelevant with a very small scaling dimension, we use a classical analogy and study the critical point of the 3d $XY$ model with a four-fold anisotropy field which is known to be irrelevant with a small scaling dimension~\cite{Hasenbusch_Vicari,Oshikawa}. Our results for the dimensionless anisotropy on this classical model are qualitatively similar to our results for the three-fold anisotropy at the N\'eel- columnar VBS transition: in both cases, the anisotropy appears to saturate to a non-zero value over the available range of sizes, although, in the classical
case, one expects it to be irrelevant at the transition; {\it (iii) } for $Q_3 \sim Q_2 \gg J$,
the competition between these two different VBS orders does {\em not} lead
to an intervening spin-liquid phase. Rather, it stabilizes
N\'eel order in a large swathe of the phase diagram even when $J$ is the smallest
energy-scale in the problem.

The article is organized as follows: in Sec. \ref{sec:modelandmethods}, we introduce the $J$-$Q_3$-$Q_2$ models that we will study and provide some computational details. In Sec. \ref{sec:pd}, we show our estimates for the phase boundaries in the $(Q_2, Q_3)$ plane. In Sec. \ref{sec:transitionline}, we study in greater detail the nature of phase-boundary $Q_{3c}(Q_2)$ separating the N\'eel phase
and the columnar VBS phase, including the behavior of the three-fold anisotropy in the phase of the columnar VBS order parameter. In Sec. \ref{sec:3dxy}, we study the classical three-dimensional
$XY$ model with four-fold anisotropy. Finally, we conclude in Sec.~\ref{sec:conc} with
a brief discussion about possible directions for future work. Some additional numerical results (on the finite-size scaling analysis of critical anisotropy, as well as 3d XY model with $q=3,5$-fold anisotropic fields) are relegated to Appendices~\ref{sec:appA} and~\ref{sec:appB}.


\section{Model and methods}
\label{sec:modelandmethods}

The main focus of our work is the numerical study of a model of spin-$1/2$ moments on sites of the honeycomb lattice, coupled
by a nearest neighbor exchange $J$ that competes with a four-spin interaction
$Q_2$ and a six-spin interaction $Q_3$:
\begin{align} 
H & =H_J + H_{Q_3} + H_{Q_2} & \\
H_J & = - J \sum_{\langle ij \rangle} P_{\langle ij \rangle} & \nonumber  \\
H_{Q_3} & = - Q_3 \sum_{\langle ij kl mn\rangle } P_{\langle ij \rangle} P_{\langle kl \rangle} P_{\langle mn \rangle} + P_{\langle jk \rangle} P_{\langle lm \rangle} P_{\langle ni \rangle}& \nonumber \\
H_{Q_2} & = - Q_2 \sum_{ \langle ij kl mn\rangle } P_{\langle ij \rangle} P_{\langle lm \rangle} + P_{\langle j k\rangle} P_{\langle mn \rangle} + P_{\langle kl\rangle} P_{\langle ni \rangle},\nonumber 
\label{eq:JQ3Q2_hamiltonian}
\end{align}
where $P_{\langle ij \rangle} = 1/4 - \mathbf{S}_i . \mathbf{S}_j$ is the singlet projector
on the bond $\langle i,j \rangle$ and $ \langle ij kl mn\rangle $ denotes
an elementary hexagon with vertices labeled cyclically (Fig.~\ref{fig:latticeandorders}c). We set $J=1$ so that all energies are measured in units of $J$.
This model is studied using the same techniques as in Ref.~\onlinecite{Pujari_Damle_Alet}, for both obtaining the ground-state and characterizing its physical properties. We summarize them here for completeness, using the same notations: we use a QMC projector algorithm~\cite{Sandvik_Evertz_PRB2010} on honeycomb lattices of linear size up to $L=60$, consisting of $L^2$ unit cells with two spins corresponding to
the two-sublattice structure of the honeycomb lattice. Periodic boundary conditions are imposed. 

N\'eel order is characterized using the vector order parameter $\vec{M} = \frac{1}{2L^2}\sum_{\vec{r}} \vec{n}(\vec{r})$,
with $\vec{n}$ the local N\'eel field $\vec{n}(\vec{r}) = \vec{S}_{\vec{r} A} - \vec{S}_{\vec{r}B}$. The unit cell is labeled by $\vec{r}$ and subscripts $A$ and $B$ refer to the two sites in this unit cell located on the different sublattices.  
The VBS order at the columnar wavevector ${\bf K} \equiv (2\pi/3,-2\pi/3)$ is characterized by the order parameter $\psi = \frac{1}{2L^2}\sum_{\vec{r}} V_{\vec{r}}$, where $V_{\vec{r}}$ is the local field:
\begin{equation*}
V_{\vec{r}} = (P_{\vec{r}0} + e^{2 \pi i /3} P_{\vec{r} 1} + e^{4 \pi i/3} P_{\vec{r} 2})e^{i {\mathbf K} \cdot \vec{r}}\; ,
\end{equation*}
with $P_{\vec{r} \mu}$ ($\mu = 0, 1,2$) the singlet projector on one of the three bonds $\mu$ corresponding to the unit cell labeled by $\vec{r}$ (see Fig.~\ref{fig:latticeandorders}). 
Finally, to quantify the staggered VBS order, we follow Ref.~\onlinecite{Banerjee_Damle_Paramekanti_2010} and use the nematic order parameter $\phi = \frac{1}{2L^2}\sum_{\vec{r}} W_{\vec{r}}$, where $W_{\vec{r}}$ is the local staggered VBS order parameter field, written as
\begin{equation*}
W_{\vec{r}} = (P_{\vec{r}0} + e^{2 \pi i /3} P_{\vec{r} 1} + e^{4 \pi i/3} P_{\vec{r} 2}) \; .
\end{equation*}
Note the absence of any $\vec{r}$ dependent phase factor in this definition. This
is consistent with the fact that staggered VBS order only breaks the symmetry
of three-fold rotations, while preserving translational symmetry.

To detect quantum phase transitions, we consider the square of the modulus of the three order parameters of interest: $\langle \vec{M}^2\rangle$, $\langle |\psi|^2\rangle = \langle \psi^{\dagger}\psi\rangle$, and $\langle |\phi|^2\rangle = \langle \phi^{\dagger} \phi \rangle$. For a continuous N\'eel-columnar VBS transition, we expect the scaling forms: $\langle \vec{M}^2\rangle = L^{-(1+\eta_{N})}f_{\vec{M}}((Q_3 - Q_{3c}^{N}) L^{1/\nu_N})$ and $\langle |\psi|^2\rangle = L^{-(1+\eta_{\rm VBS})}f_{\psi}((Q_3 - Q_{3c}^{D}) L^{1/\nu_D})$. In writing these scaling forms, we assume
that the phase boundary is crossed by varying $Q_3$ at fixed $Q_2$ and allow
for two different correlation length exponents $\nu_{N/D}$ associated with N\'eel / columnar VBS correlations at different critical values $Q_{3c}^{N/D}$. We do not quote the scaling form for the staggered VBS order as this transition is strongly first order. 

We also use the following Binder ratios $g_{{M}} = {\langle (\vec{M}^2)^2\rangle} / {\langle \vec{M}^2\rangle^2}$, $g_{\psi} = \langle | \bar{E_{\psi}} |^4 \rangle/\left( \langle |\bar{E_{\psi}}|^2\rangle \right)^2$ and $g_{\phi} = \langle | \bar{E_{\phi}} |^4 \rangle/\left( \langle |\bar{E_{\phi}}|^2\rangle \right)^2$ to locate the quantum critical points where N\'eel, columnar and staggered VBS orders respectively disappear. The two first Binder ratios are expected to scale close to a continuous quantum phase transitions as $g_{{M}} = g_{{M}}((Q_3 - Q_{3c}^{N}) L^{1/\nu_N})$ and $g_{\psi} = g_{\psi}((Q_3 - Q_{3c}^{D}) L^{1/\nu_D})$ respectively. Note that both VBS Binder ratios are not written in terms of the powers of the corresponding VBS order parameter, as this would involve computations of $8-$spin correlation functions, for which there is no simple expression in the valence-bond formalism used in the QMC simulations. Instead, we use moments of the Monte-Carlo estimator  $E_{\psi}$\cite{Sandvik_PRL2007,Lou_etal_PRB09} (respectively $E_{\phi}$), whose Monte-Carlo average $\bar{E_{\psi}}$ (respectively $\bar{E_{\phi}}$) coincides with the quantum-mechanical expectation value $\langle \psi \rangle$ ($\langle \phi \rangle$) of the columnar (resp. staggered) VBS order parameter. In all our simulations, we found that this correctly reproduces the expected physical behavior for moments of $\psi $ or $\phi$.
 
Close to continuous quantum phase transitions, we have fitted our numerical data to the respective scaling forms, using polynomial up to second order in most cases for the universal functions $f_{M / \psi}$ and $g_{M / \psi}$.

We now introduce the observables related to the phase of the columnar VBS order parameter $\psi$. The phase of $\psi$ distinguishes a fixed columnar (`Kekul\'e') pattern of bond-energy expectation values from one in which a sublattice of plaquettes hosts
a valence-bond resonance (see Fig.~\ref{fig:latticeandorders}). Both patterns
correspond to a three-fold symmetry breaking
and lead to order at the same wavevector ${\mathbf K}$, but they differ in the phase
of the complex VBS order parameter $\psi$. In our QMC simulations, we do not have access strictly speaking to the phase of $\psi$, but rather to the phase $\theta_{E_{\psi}}$ of the estimator $E_{\psi} \equiv |E_{\psi}| \exp(i\theta_{E_{\psi}})$. We nevertheless expect that it reflects the behavior of the true phase of $\psi$.
To address the relevance of $3$-fold monopole events, we consider the following dimensionless measure of the anisotropy in the distribution of this phase: 
\begin{eqnarray}
W_3=\int d E_{\psi} P(E_{\psi}) \cos(3\theta_{E_{\psi}}) 
\end{eqnarray}
with $P(E_{\psi})$ is the normalized probability distribution for this quantity as sampled
by the Monte-Carlo run.

It is also possible to analyze our data using scaling theories~\cite{Oshikawa,Lou_Sandvik_Balents,okubo} to capture the finite-size behavior of $W_3$ near criticality. We have used such a scaling analysis to fit our numerical data as detailed in Appendix A, but we prefer to display the bare numerical data for the anisotropy measure $W_3$ in Sec.~\ref{sec:3fold} in order to avoid any assumption regarding the scaling form obeyed by $W_3$.

\section{Phase diagram}
\label{sec:pd}

\begin{figure}
\includegraphics[width=0.8 \hsize]{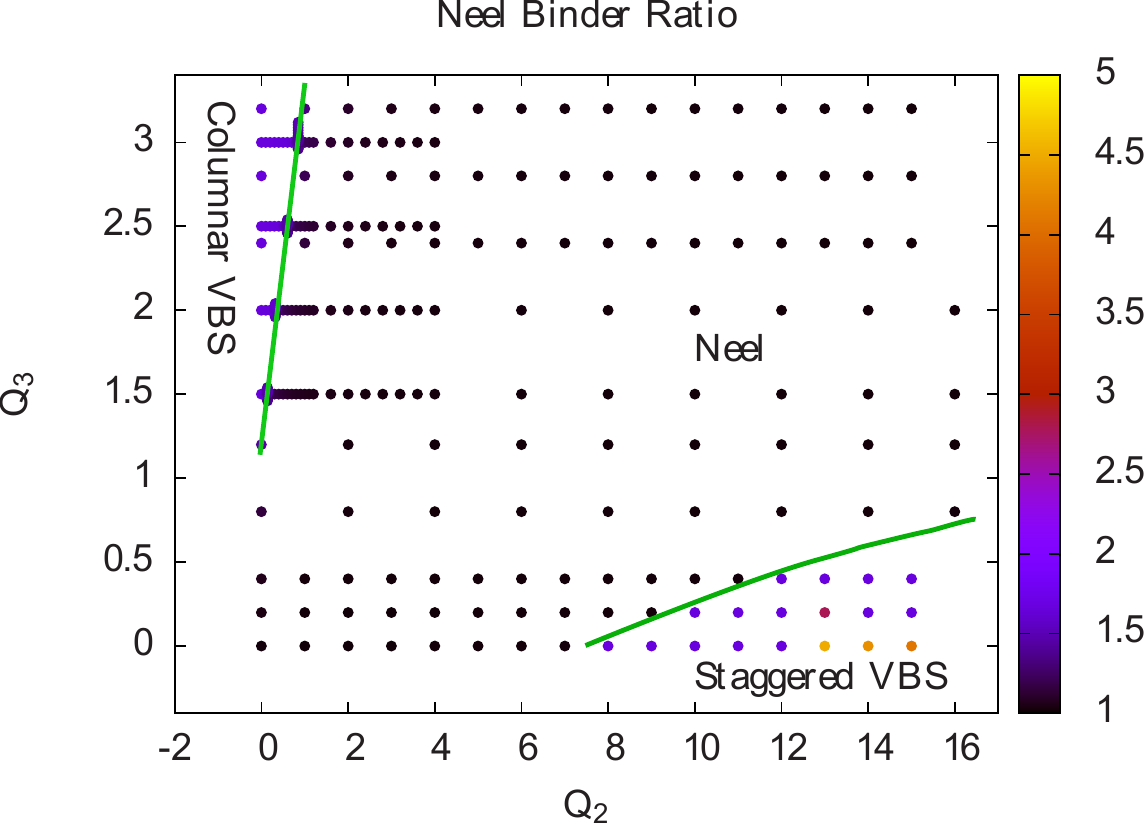}
\\ \vspace{2mm}
\includegraphics[width=0.8 \hsize]{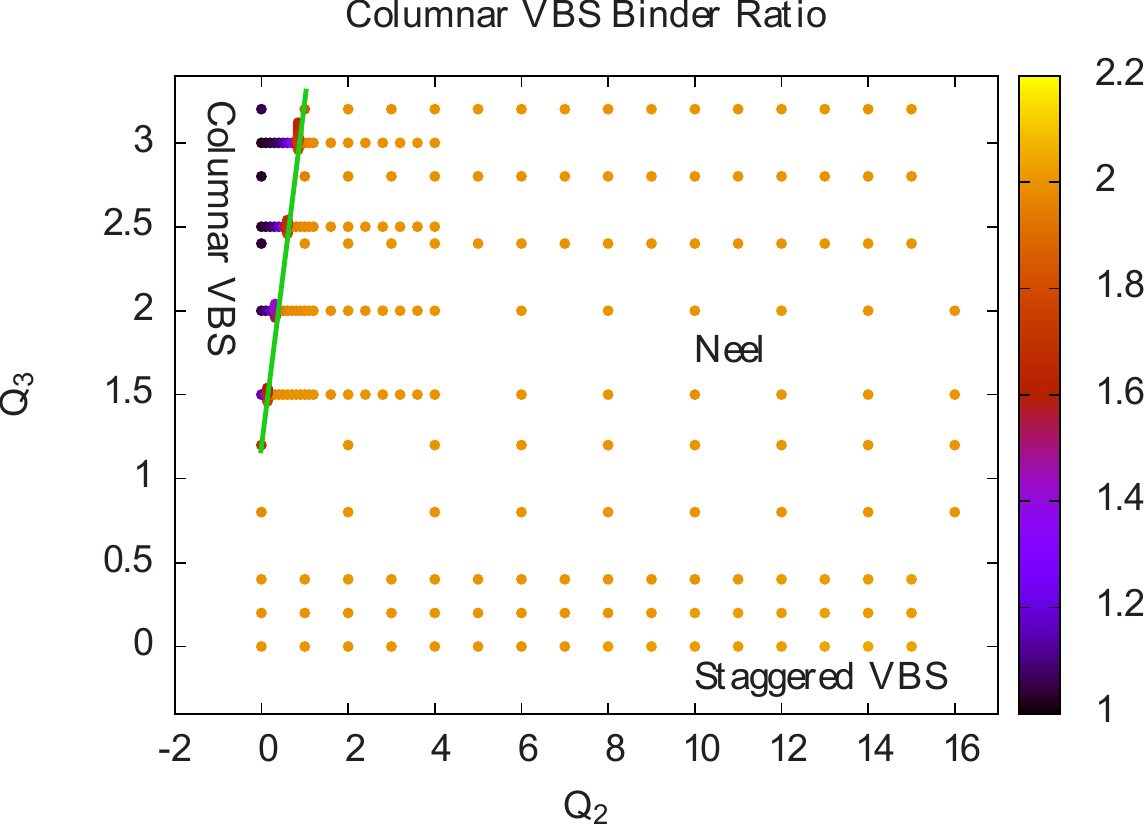}
\\ \vspace{2mm}
\includegraphics[width=0.8 \hsize]{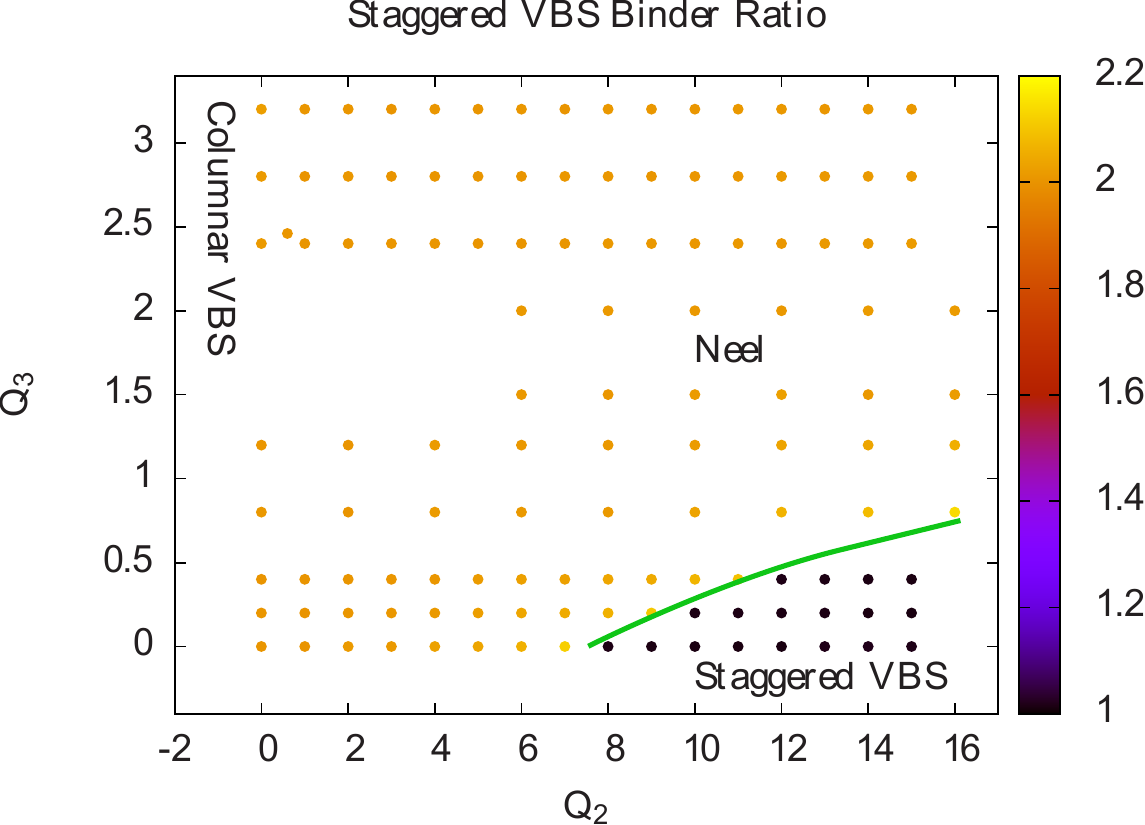}
\caption{(Color online) Color maps of Binder cumulants (top : N\'eel Binder cumulant $g_M$, middle: VBS columnar Binder cumulant $g_\psi$, bottom: staggered Binder cumulant $g_\phi$) for different values in the ($Q_2,Q_3$) parameter space, for a system of linear size $L=24$. Low values indicate long-range order, while high values indicate absence of order. These results allow to map the phase diagram where N\'eel, columnar and staggered VBS phases can be identified (green lines are indications of approximate phase boundaries). }
\label{fig:allBinder}
\end{figure}

We first present our results on the phase diagram of the ground-state of $H$ in the $(Q_2,Q_3)$ plane. 
As noted earlier, $H_J$ favors N\'eel ordering, while $H_{Q_2}$ ($H_{Q_3}$) favor staggered
(columnar) VBS order. We can locate two limiting points using results from previous works. The model $H_J+H_{Q_3}$ has been shown~\cite{Pujari_Damle_Alet} to host a continuous phase transition from the N\'eel to a columnar VBS state at $Q_{3c}(Q_2=0) \simeq 1.19$. Since $Q_2$ disfavors columnar VBS order, we expect the phase
boundary $Q_{3c}(Q_2)$ between the N\'eel state and the columnar VBS state
to define an increasing function of $Q_2$, at least for small $Q_2$. On the other hand, Ref~\onlinecite{Banerjee_Damle_Paramekanti_2010} showed that the model $H_{J}+H_{Q_2}$ exhibits a strongly first-order transition 
from the N\'eel to the staggered VBS state at $Q_{2c}(Q_3=0) \simeq 6.4$. We expect the first-order transition to staggered VBS order to shift to increasing values of $Q_2$ when $Q_3$ is turned on. 

A first estimate on the location of these phase boundaries is given by the magnitude of the N\'eel Binder cumulant $g_M$. In our definition of $g_M$, and for a large enough system size, a value close to $1$ corresponds to a phase with antiferromagnetic order, while a value $5/3$ corresponds to gaussian fluctuations centered at zero, signaling no magnetic order. At the quantum N\'eel-columnar VBS critical point at $Q_2=0$, the N\'eel Binder cumulant takes~\cite{Pujari_Damle_Alet} a value $\simeq 1.42$ (which should be universal), lying between these two limiting values. In contrast, close to a first-order transition~\cite{vollmayr}, this Binder cumulant can take values larger than $5/3$ on finite-systems. We display
the magnitude of $g_M$ for a system of moderate size $L=24$ in the top panel of Fig.~\ref{fig:allBinder}.
This allows a first estimate of the phase boundaries: we clearly observe two transition lines emerging from the limiting points at $Q_3=0$ and $Q_2=0$. The nature of the transitions does not appear to change, since we observe very high values for $g_M$ (signaling a first-order transition) for the line emerging from $Q_2^c(Q_3=0)$, and intermediate values (between $1$ and $5/3$) for the line emerging from $Q_3^c(Q_2=0)$, signaling a continuous transition. This is confirmed by a finite-size scaling analysis in the next section. 
From this study of $g_M$, we also see that antiferromagnetism survives in the region $Q_2,Q_3 \gg J$. Thus, the competition between the two VBS orders does not lead
to spin-liquid behavior. Rather, it allows antiferromagnetism to set in although
$J$ is the smallest energy scale in the Hamiltonian.
The phases where no antiferromagnetism is present are naturally expected to host columnar (at low $Q_2$) and staggered (low $Q_3$) VBS orders. This is well confirmed by the low values (close to 1) taken by the columnar and staggered VBS Binder cumulants displayed in the middle and bottom panels of Fig.~\ref{fig:allBinder}.

\begin{figure}
\includegraphics[width=\hsize]{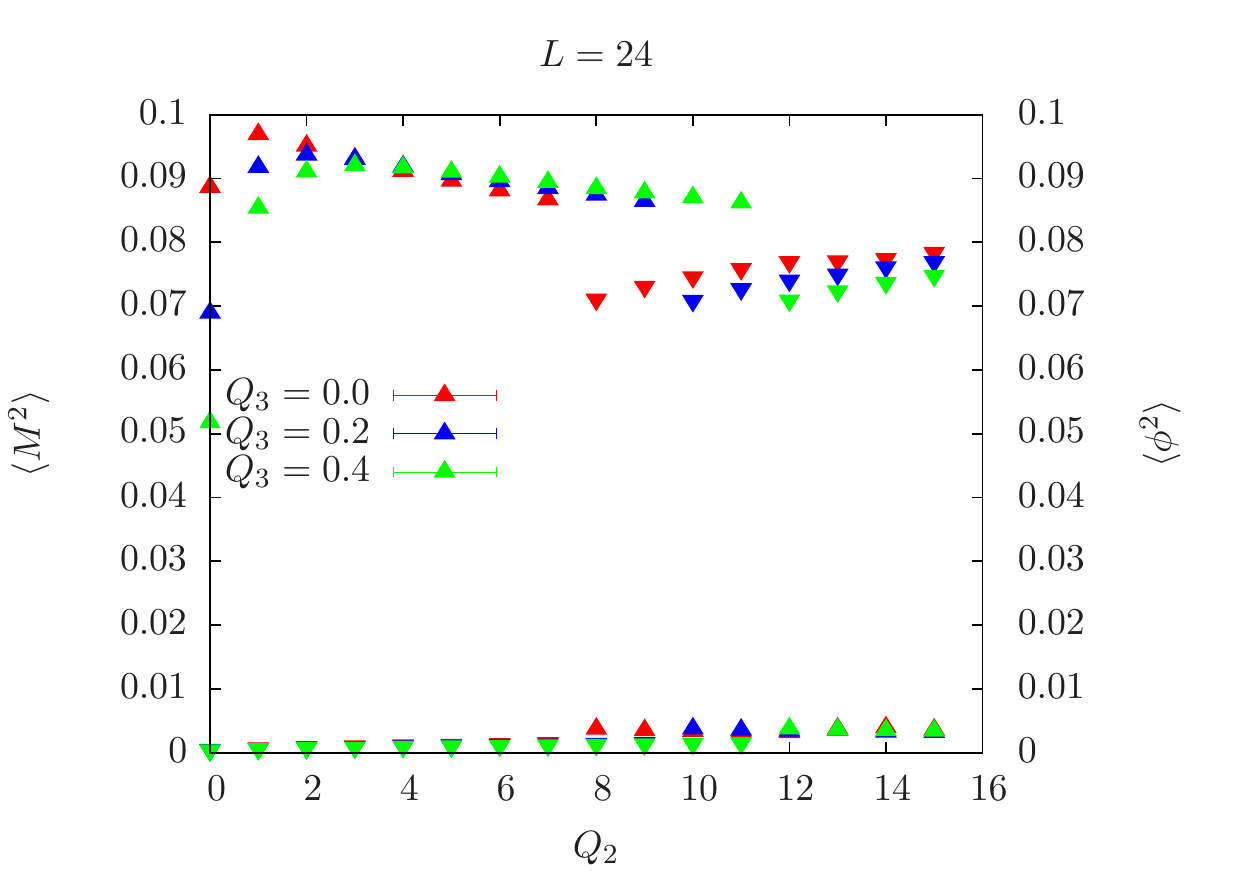}
\caption{The first-order transition from N\'eel to staggered VBS order
is readily identified by the sharps jumps in the corresponding order parameters (up triangles: $\langle M^2\rangle$, down triangles $\langle \phi^2 \rangle$), for three values of $Q_3$ (system size $L=24$).}
\label{fig:orderparameterjumps}
\end{figure}

We now consider more carefully the transition line $Q_{2c}(Q_3)$ between the N\'eel and staggered VBS order, by locating the abrupt first-order jumps in the two order parameters. An example of these
jumps is shown in Fig.~\ref{fig:orderparameterjumps} and the resulting phase boundary is represented as a line in Fig.~\ref{fig:allBinder}. 
The transition between the N\'eel and columnar VBS transitions deserves a more careful finite-size scaling analysis, which is presented in Sec.~\ref{sec:transitionline}: the resulting transition line $Q_{3c}(Q_2)$ is also represented in Fig.~\ref{fig:allBinder}.

\section{N\'eel-columnar VBS transition line}
\label{sec:transitionline}

\subsection{Exponents and scaling forms}
\label{subsec:exponents_and_scaling_forms}

We focus here on the nature of the phase-boundary between the N\'eel and
the columnar VBS states. Following our earlier work~\cite{Pujari_Damle_Alet}
at $Q_2=0$, we locate the point at which N\'eel order is lost using the dimensionless
Binder ratio $g_{{M}}$, and the point at which the columnar VBS order turns
on using the corresponding Binder ratio $g_{\psi}$. For four different values of $Q_2$, we vary $Q_3$ to locate the quantum phase transition and attempt to collapse the Binder ratio data onto the corresponding scaling forms (see Sec.~\ref{sec:modelandmethods}). In the analysis, we allow these two scaling forms to use different
values of $Q_{3c}$ as well as different correlation length exponents $\nu_N$
and $\nu_D$. We also analyze the collapse of the modulus squares of order parameters $\langle M^2 \rangle$ and $\langle |\psi^2| \rangle$ according to the forms in Sec.~\ref{sec:modelandmethods}, providing estimates of $Q_{3c},\nu_N,\nu_D$ as well as $\eta_N$ and $\eta_D$. 

\begin{figure}[h]
\includegraphics[width=0.8 \hsize,angle=0]{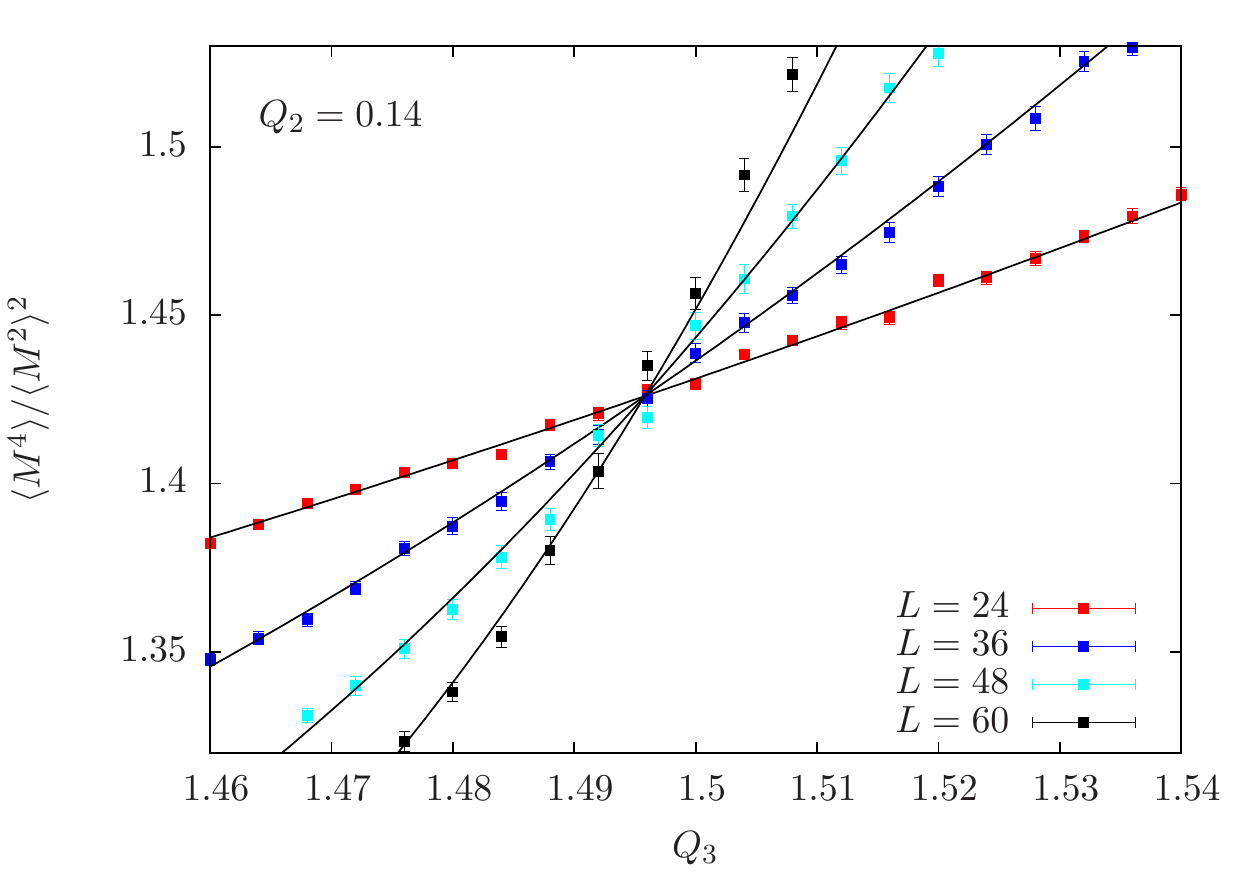}
\includegraphics[width=0.8 \hsize,angle=0]{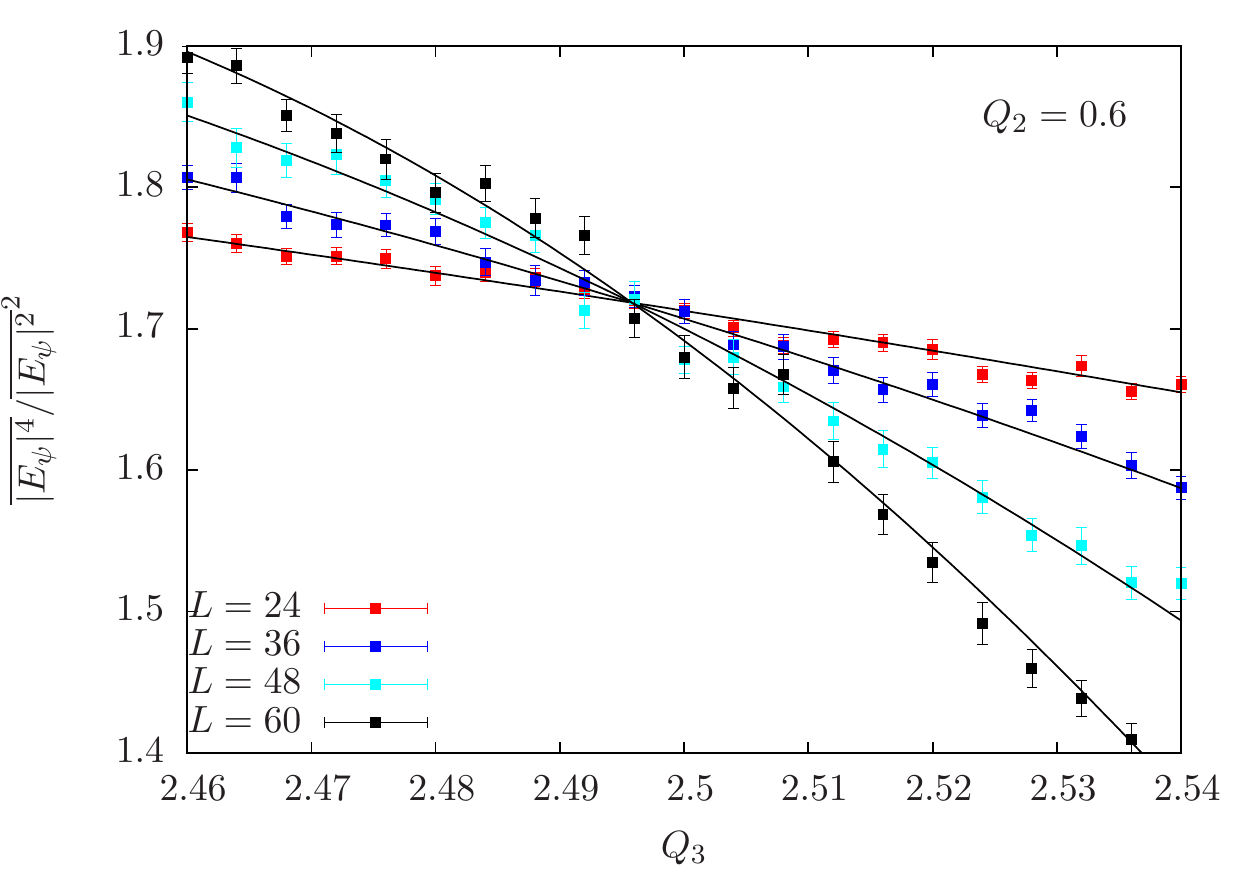}
\caption{(Color online) Crossing plot of Binder cumulants for different system sizes : N\'eel  cumulant $g_{{M}}$ (top panel, for $Q_2=0.14$) and columnar VBS  cumulant $g_{\psi}$ (bottom panel, for $Q_2=0.60$). Symbols are QMC data, solid lines fits to the finite-size scaling form (see text). 
For the fits, a particular choice of critical window, minimum system size included, and 
order of universal function has been shown here which gave $\chi^2$ per degree of freedom
equal to 1.53
and 0.97 for the plots respectively. For estimates
on overall error-bars, refer to Table \ref{table1}.
  \label{fig:Binder}
}
\end{figure}

\begin{figure}[h]
\includegraphics[width=0.8 \hsize,angle=0]{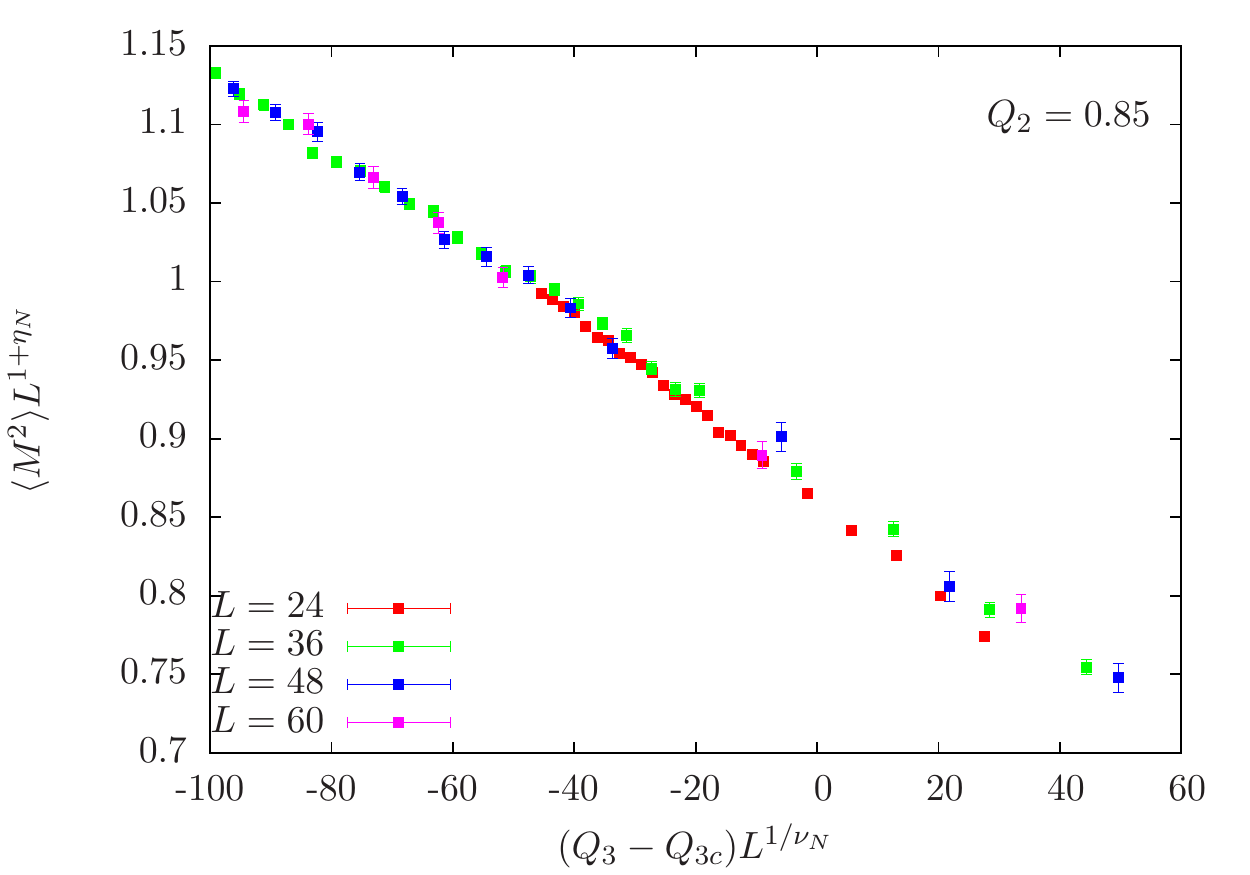} \\
\includegraphics[width=0.8 \hsize,angle=0]{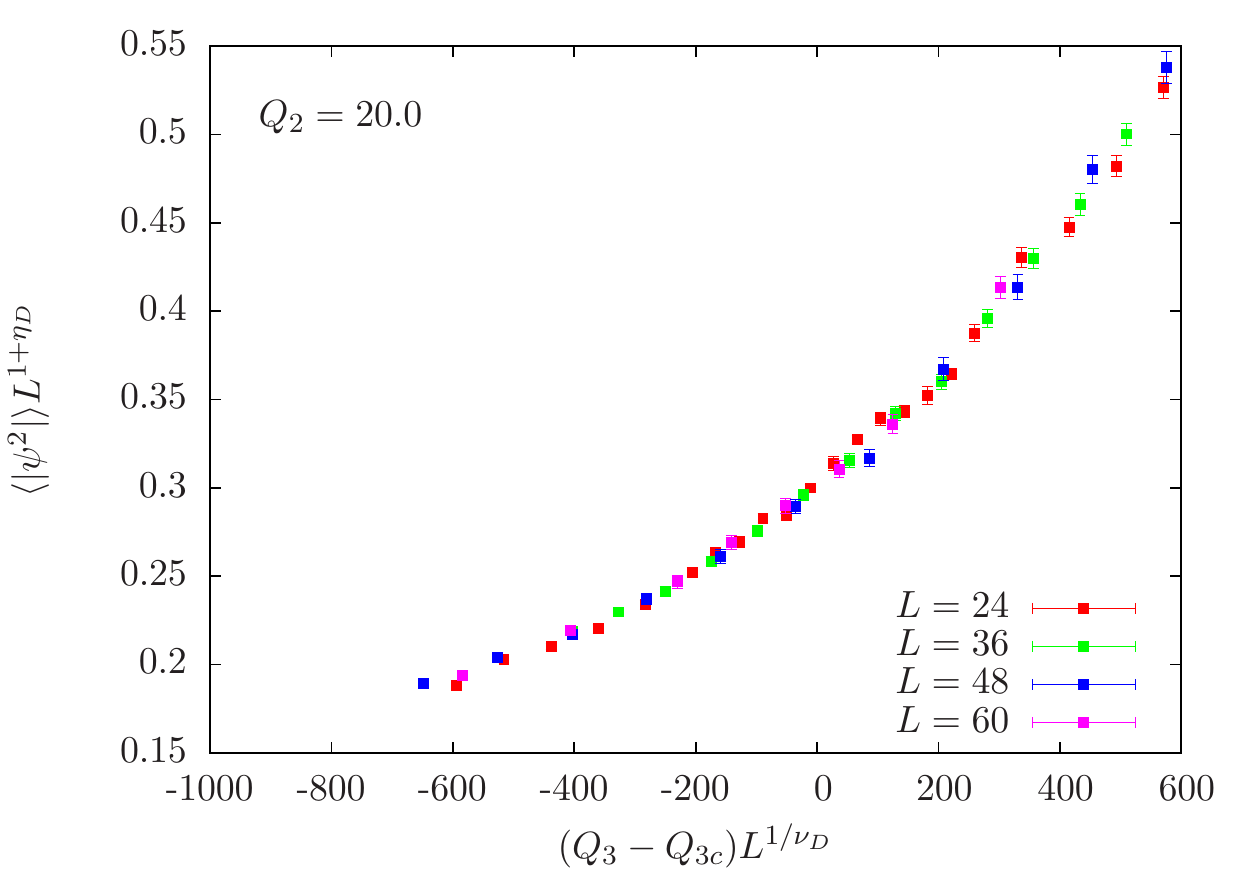}
\caption{(Color online) Scaling collapse for different system sizes of the N\'eel order parameter $\langle {M}^2 \rangle$ (top panel, $Q_2=0.85$) and columnar VBS order parameter  $\langle |\psi|^2 \rangle$ (bottom panel, $Q_2=20.0$) in the critical region. Critical point $Q_{3c}$ and critical exponents are obtained by fits to the standard finite-size scaling forms (see text). 
For the fits, again a particular choice of critical window, minimum system size included, and 
order of universal function has been shown here which gave $\chi^2$ per degree of freedom
equal to 1.25
and 1.15 for the plots respectively. For estimates
on overall error-bars, refer to Table \ref{table1}.
\label{fig:OP}
}
\end{figure}

In Figs. \ref{fig:Binder} and \ref{fig:OP}, we provide representative
examples of the results of such an analysis. Our data all along the N\'eel-columnar VBS phase boundary is well-described by conventional scaling forms. 
For ready-reference, we also tabulate estimates of the corresponding critical points, exponents and amplitudes  values obtained using these different observables in Table~\ref{table1}.

\begin{table*}
     \begin{ruledtabular}

\begin{tabular}{|c|c|c|c|c|c|c|c|c|c|c|c|c|}
\rule[-1.2ex]{0pt}{0pt}   & \multicolumn{3}{c|}{$\langle M^2 \rangle$} & \multicolumn{3}{c|}{$g_M=\langle M^4 \rangle / \langle M^2 \rangle^2$} & \multicolumn{3}{c|}{$\langle |\psi|^2 \rangle$} & \multicolumn{3}{c|}{$g_\psi=\langle \overline{|E_\psi|^4} \rangle / \langle \overline{|E_\psi|^2} \rangle^2$} \\
    
    \hline
$Q_2$ & $Q_{3c}$ & $\nu_N$  & $\eta_N$ & $Q_{3c}$ & $\nu_N$  & $g_M(0)$ & $Q_{3c}$ & $\nu_D$  & $\eta_D$ & $Q_{3c}$ & $\nu_D$  & $g_\psi(0)$ \\
\hline
0.14 & 1.496(2) & 0.58(2) & 0.27(3) & 1.496(1) & 0.57(3) & 1.425(2) & 1.483(2) & 0.59(2) & 0.37(3) & 1.491(1) & 0.57(3) & 1.718(5) \\
0.60 & 2.506(2) & 0.56(2) & 0.31(2) & 2.500(1) & 0.56(2) & 1.427(1) & 2.491(5) & 0.57(3) & 0.23(7) & 2.495(1) & 0.56(2) & 1.721(3) \\
0.85 & 3.058(2) & 0.55(4) & 0.33(2) & 3.050(2) & 0.56(2) & 1.428(3) & 3.03(1) & 0.60(3) & 0.26(8) & 3.044(2) & 0.56(2) & 1.721(5)\\
20.0 & 45.3(1) & 0.57(2) & 0.31(3) & 45.27(2) & 0.56(2) & 1.430(2) & 45.0(1) & 0.61(3) & 0.32(6) & 45.18(1) & 0.56(2) & 1.727(1)\\
\end{tabular}
     \end{ruledtabular}
\caption{ \label{table1}
For different values of $Q_2$ : estimates of critical point, exponent and amplitudes resulting from the finite-size scaling analysis of order parameters $\langle M^2 \rangle$, $\langle |\psi|^2 \rangle$ and associated Binder cumulants $\langle M^4 \rangle / \langle M^2 \rangle^2$, $\langle \overline{|E_\psi|^4} \rangle / \langle \overline{|E_\psi|^2} \rangle^2$.
Error bars were determined from the spread on extracted fit parameters  depending
on critical window size, minimum system sizes included, 
and degree of polynomial for the universal scaling functions, with $\chi^2$ per degree of freedom always
$\lesssim 1.5$.}
\end{table*}

We find that these estimates of $Q_{3c}$ at a given value
of $Q_2$ agree approximately with each other within statistical errors.
More precisely, the spread in the best-fit values of $Q_{3c}$ obtained from VBS data in two
different ways (from $g_{\psi}$ and  $\langle |\psi^2| \rangle$) is of the same order as 
the difference in the best-fit $Q_{3c}$ values obtained from scaling collapses of $g_{\psi}$  and $g_M$. The same is true
for the correlation length exponents $\nu_N$ and $\nu_D$ at a given value of $Q_2$. Therefore, we conclude that one can consistently account for all the data at a given value of $Q_2$ in terms of a single critical point $Q_{3c}(Q_2)$ at which N\'eel order is lost and columnar
VBS order turns on, with both N\'eel and columnar order
parameters controlled by a single correlation length exponent $\nu$. Within errors, this
estimate of $\nu$ does not exhibit any $Q_2$ dependence. The anomalous exponents $\eta_N$ and $\eta_D$ are also found to be $Q_2$-independent within error bars (which are larger for $\eta_D$). Additionally, we note that $\eta_N$ and $\eta_D$ are close to each other in value (although the theory of deconfined criticality does not predict that these anomalous
dimensions are equal). The amplitudes of both VBS and Binder ratios at criticality are also found to be constant within errors along the critical line. Finally, we emphasize that all estimates of the critical exponents and amplitudes for $Q_2\neq 0$ agree with those found in the case $Q_2=0$~\cite{Pujari_Damle_Alet}.

Our numerical simulations therefore indicate that the entire N\'eel-columnar VBS transition line belongs to a single universality class. Our estimates for the critical exponents are very
close to the latest estimates for SU($2$) models on the square lattice~\cite{Sandvik_2012,Block_Melko_Kaul,Harada_Suzuki_etal} suggesting that both honeycomb and square lattice
transitions are in the same universality class, presumably described by the NCCP$^1$ critical theory.
This strongly suggests that three-fold monopole events are irrelevant at the N\'eel-columnar VBS critical point for a SU($2$) model on the honeycomb lattice.

\subsection{Three-fold anisotropy at criticality}
\label{sec:3fold}

Given that the entire phase boundary appears to be controlled by a single
fixed point, it is of interest to investigate the $Q_2$ dependence of the three-fold
anisotropy in the phase of the columnar VBS order parameter $\psi$ at criticality.
To this end, we focus on the histogram of $E_{\psi}$ measured at and in the close
vicinity of our best estimate for $Q_{3c}(Q_2)$.
The simplest methodology is one that requires the fewest theoretical assumptions
about the scaling properties of the
three-fold anisotropy. In this approach, we simply monitor the large-$L$ behavior of the dimensionless anisotropy
measure $W_3$ (as defined in Sec.~\ref{sec:modelandmethods}) for a few values
in the vicinity of $Q_{3c}(Q_2)$ for various values of $Q_2$. This $L$ dependence is interpreted by noting that  $W_3$ tends to zero (respectively to unity) with increasing system size deep in the N\'eel (resp. columnar VBS) phase. If three-fold anisotropy is irrelevant at the transition, one would expect $W_3$ to tend to zero for large $L$ at
the critical point, but increase with increasing $L$ when one moves into the VBS phase.

\begin{figure}[h!]
\includegraphics[width=0.8 \hsize,angle=0]{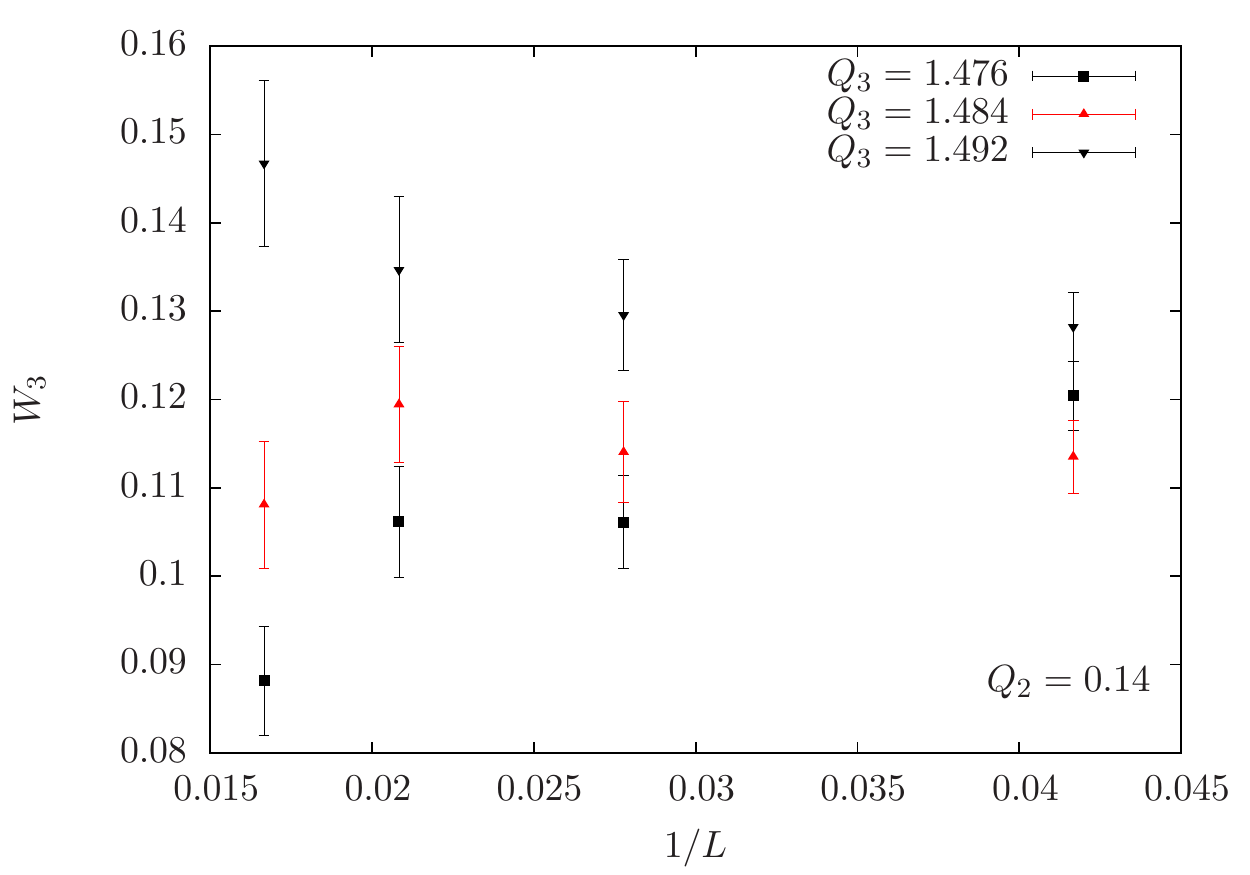}
\includegraphics[width=0.8 \hsize,angle=0]{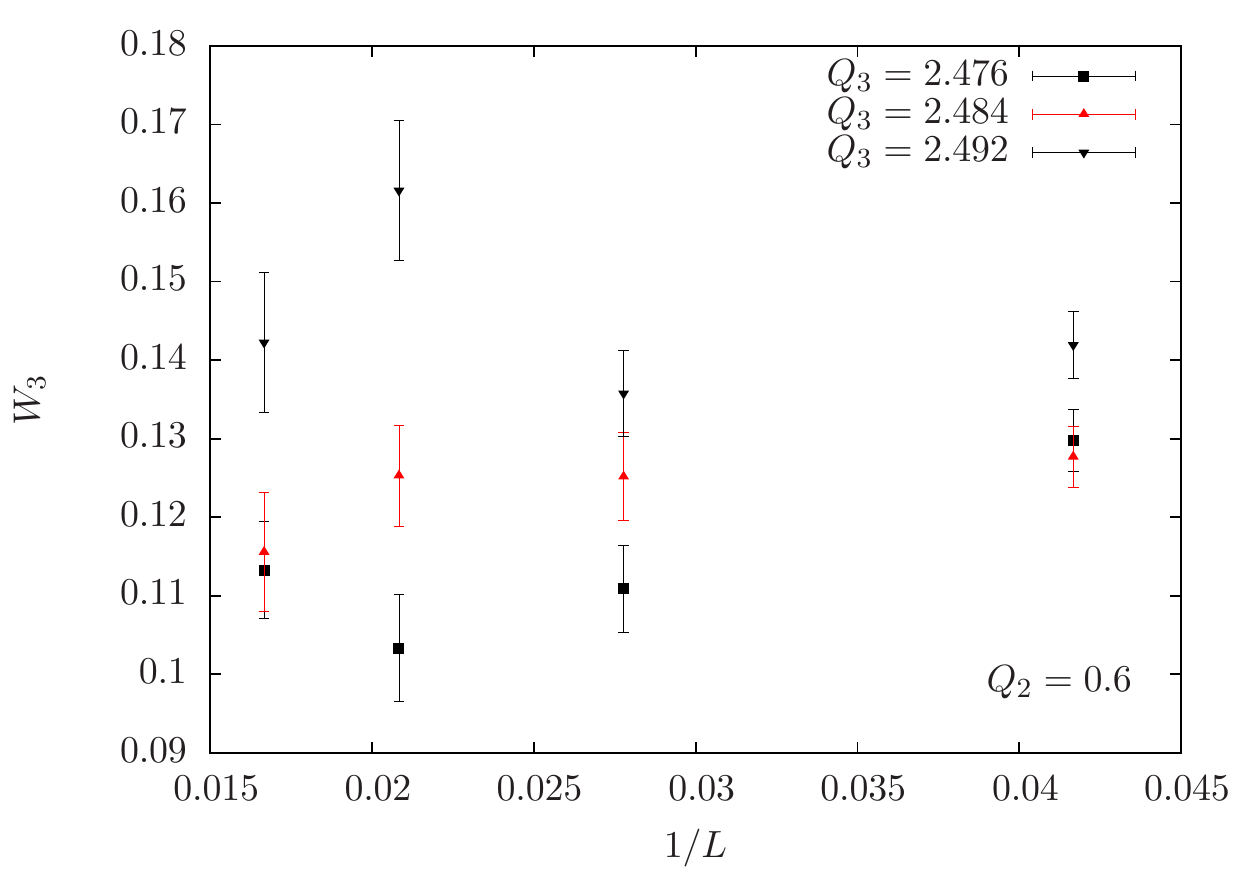}
\includegraphics[width=0.8 \hsize,angle=0]{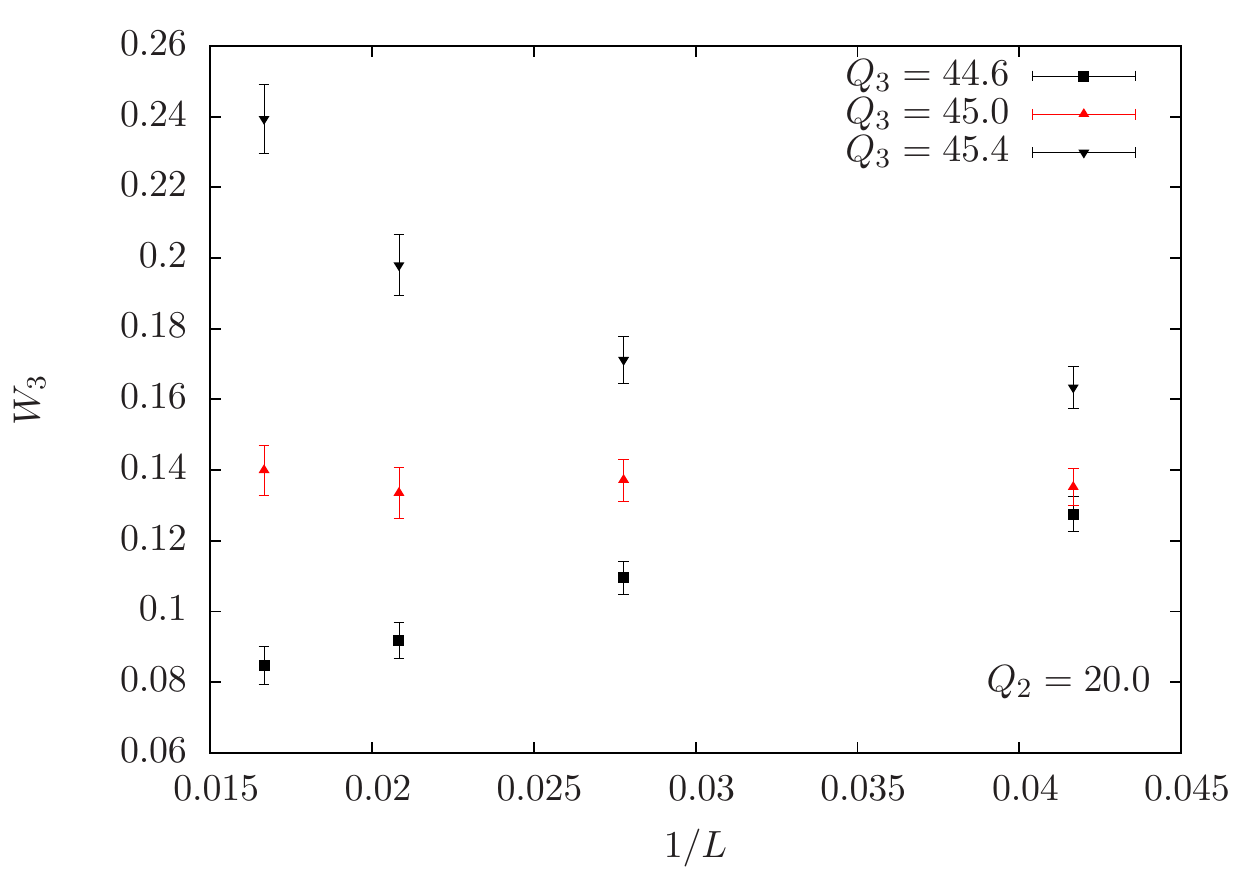}
\caption{(Color online) Finite-size dependence of $W_3$  close to the critical point for $Q_2=0.14$ (top panel), $Q_2=0.60$ (middle panel), $Q_3=20.0$ (bottom panel). In each case, we display data for one value of $Q_3$ closer to our  estimate of $Q_{3c}$, one slightly above and one slightly below.
\label{fig:W3}
}
\end{figure}

In Fig. \ref{fig:W3},  we display the $L$ dependence of this quantity
in the vicinity of $Q_{3c}(Q_2)$ for three different values of $Q_2$, two small and one large. From this data, it is clear that our earlier finding\cite{Pujari_Damle_Alet}, of an apparently non-zero large-$L$ limit for this quantity at criticality, remains valid all along
the N\'eel-VBS phase boundary, including at the largest value of $Q_2$ studied.
This nonzero limiting value $W_{3c}$ appears to increase slightly with $Q_2$, as can already be observed in Fig.~\ref{fig:W3}. 
A critical window around $W_{3c}$ can be defined by considering the values taken by this dimensionless anisotropy in the critical region around $Q_{3c}$ obtained from the analysis of the previous section. In this window, one can attempt a more sophisticated scaling analysis
that uses some assumptions about the structure of the scaling theory for $W_3$. This is
presented in Appendix A, and provides independent estimates of $W_{3c}$ from fits to
a scaling form. These estimates, and the resulting conclusions are consistent with those presented above from the more direct analysis above.

We are thus led to two conclusions that appear, at first sight, to contradict each other.
The first is that critical exponents and values of Binder cumulants at criticality along the entire phase boundary are compatible with the NCCP$^1$ universality class. The second
is that this is accompanied by a non-vanishing three-fold anisotropy of the phase
of $\psi$ at criticality, which furthermore appears to vary (albeit slightly) along the critical
line. As we show in the next section, in the better-understood
classical example of a 3d $XY$ model with weakly-irrelevant
four-fold anisotropy, the dimensionless anisotropy at criticality again appears
to saturate to a non-zero large-$L$ limit when studied over a limited range
of sizes accessible to Monte-Carlo simulations. As argued in the next
section, this suggests a possible
rationalization of our findings: three-fold anisotropy is indeed irrelevant
at the N\'eel-columnar VBS transition, but only very weakly so.

\section{Classical $3d-XY$ model with $Z_4$ anisotropy on the cubic lattice}
\label{sec:3dxy}

We find it useful to compare this peculiar, apparently non-zero large $L$ limit
of $W_3$ at criticality to the behavior of an analogous quantity in a much simpler
classical setting in which one can explicitly tune the bare value of the corresponding
anisotropy, namely the $3d-XY$ model with $Z_4$ anisotropy on the cubic lattice.
This choice of analogy is dictated by the following considerations: from earlier
work, we know that
$Z_3$ anisotropy is relevant at the isotropic $3d-XY$ transition, driving
the system to a weakly first-order transition, while $Z_4$ and higher anisotropies
have all been found to be irrelevant at the isotropic $XY$ transition (with $Z_4$
anisotropy having the smallest scaling dimension among the irrelevant terms). 
These conclusions are based on an $\epsilon$-expansion of the corresponding field theory~\cite{Oshikawa}, Monte Carlo estimates of the scaling dimensions of $q-$fold anisotropy terms~\cite{Hasenbusch_Vicari}, as well as direct numerical simulations of the $3d-XY$ model with $Z_{q\geq 4}$ anisotropies (as e.g. in Ref.~\onlinecite{Lou_Sandvik_Balents}) and of the $q=3$ states Potts model~\cite{Janke}.

Thus, by adding a $Z_4$ anisotropy field $h_4$ to the isotropic $3d-XY$ model and
studying the critical point as a function of $h_4$, we can study an example
of critical behavior in the presence of an irrelevant anisotropy which scales
to zero very slowly (since it has a small scaling dimension). This provides
us a setting to explore via analogy the possibility that
the nonzero $W_{3c}$ observed for all $Q_2$ along the N\'eel-columnar VBS
phase boundary could reflect the fact that three-fold anisotropy is irrelevant at this
transition, but has small enough scaling dimension that it appears almost marginal
(saturating to a non-zero value) in the range of sizes accessible to numerics.

We consider the 3d classical ferromagnetic $XY$ model with a $Z_4$ anisotropy term, defined by the Hamiltonian 
\begin{equation}
\mathcal{H} = - \sum_{\langle \vec{r}, \vec{r}'\rangle}  \cos(\theta_{\vec{r}} - \theta_{\vec{r}'}) 
    - h_4 \sum_{\vec{r}}\cos(4 \theta_{\vec{r}})
\label{eq:3dxy_hamiltonian}
\end{equation}
where $\langle \vec{r},\vec{r}'\rangle$ denotes nearest-neighbor sites on the simple cubic lattice and $\theta_{\vec{r}}$ are $U(1)$ angular variables $\in [ 0, 2 \pi ) $ at site ${\vec{r}}$. This model has a high-temperature paramagnetic
phase where the $U(1)$ symmetry is unbroken, and a low temperature ordered phase where the spins align in one of the $4$ preferred directions. At $h=0$, the model has a $U(1)$ symmetry which is spontaneously broken in the low-temperature phase. To access this physics, we perform classical Monte Carlo simulations on simple cubic lattice of linear sizes $L\in\{8,16,24,32,48,64\}$ with periodic boundary conditions using a combination of local Metropolis and Wolff cluster updates\cite{Wolff}. 

\begin{figure}
\includegraphics[width=0.8 \hsize,angle=0]{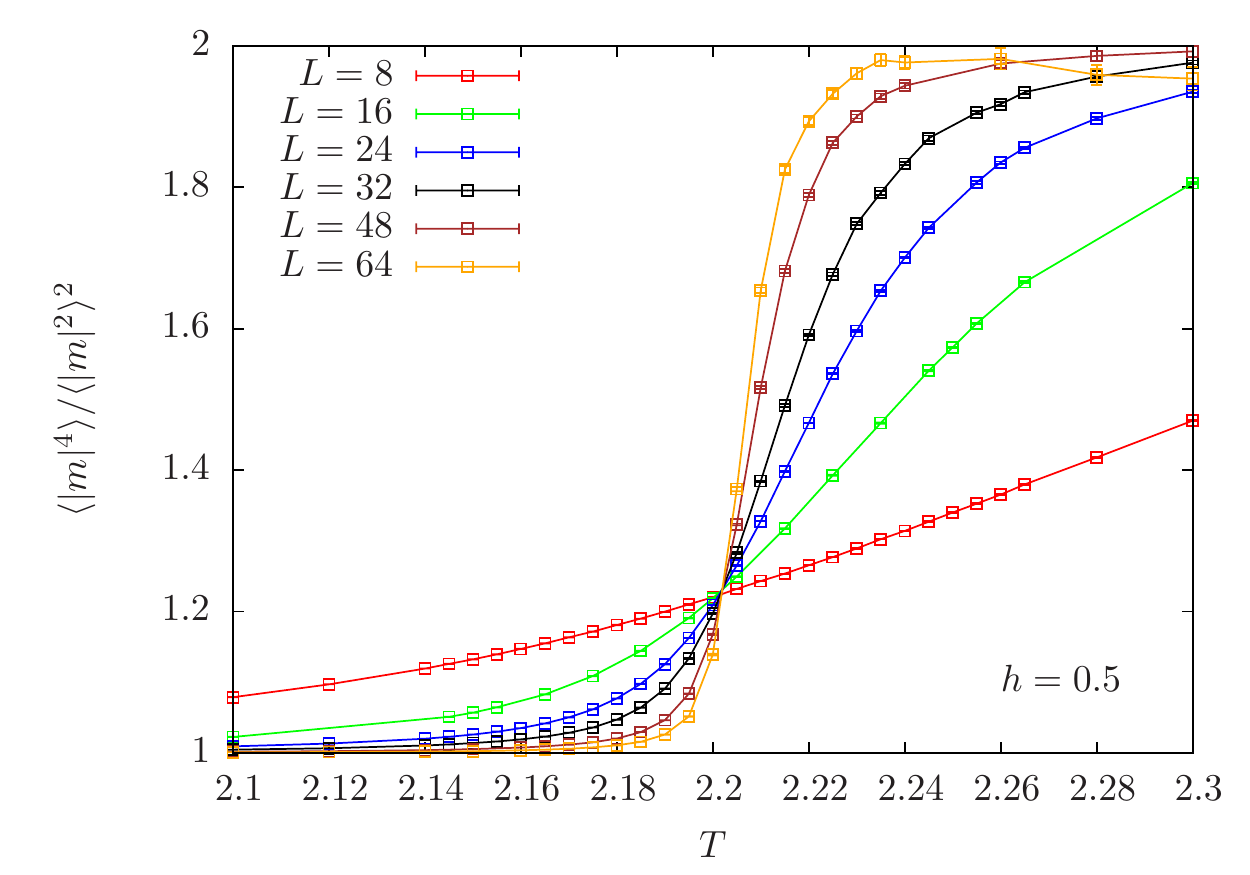}
\includegraphics[width=0.8 \hsize,angle=0]{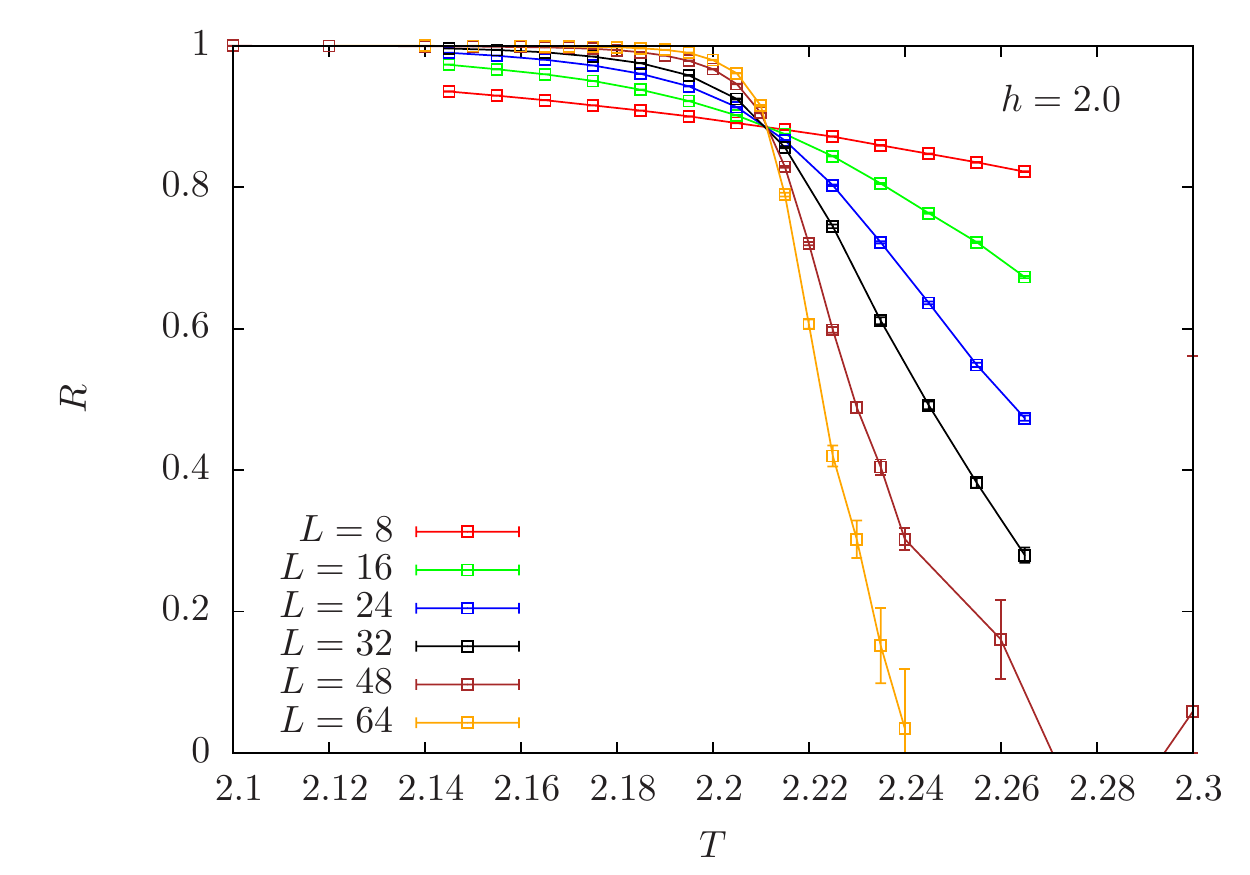}
\vspace{-2.5mm}
\caption{(Color online) 3d XY model with four-fold anisotropic field: crossing plot for different system sizes for the Binder cumulant $B$ (top panel, for $h=0.05$) and correlation ratio $R$ (bottom panel crossing plot, for $h=2$).
\label{fig:3dxy_BR}
}
\end{figure}

\begin{figure}
\includegraphics[width=0.8 \hsize,angle=0]{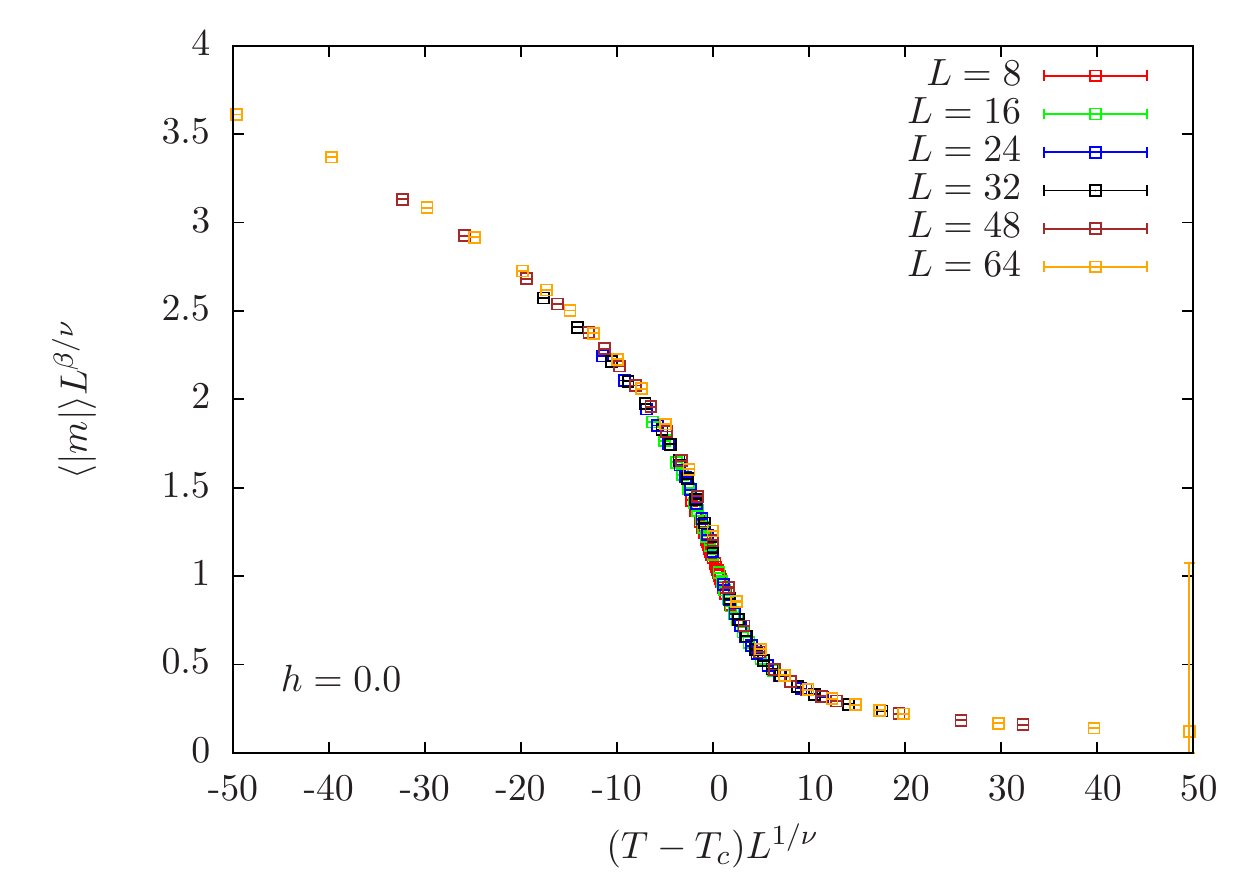}
\includegraphics[width=0.8 \hsize,angle=0]{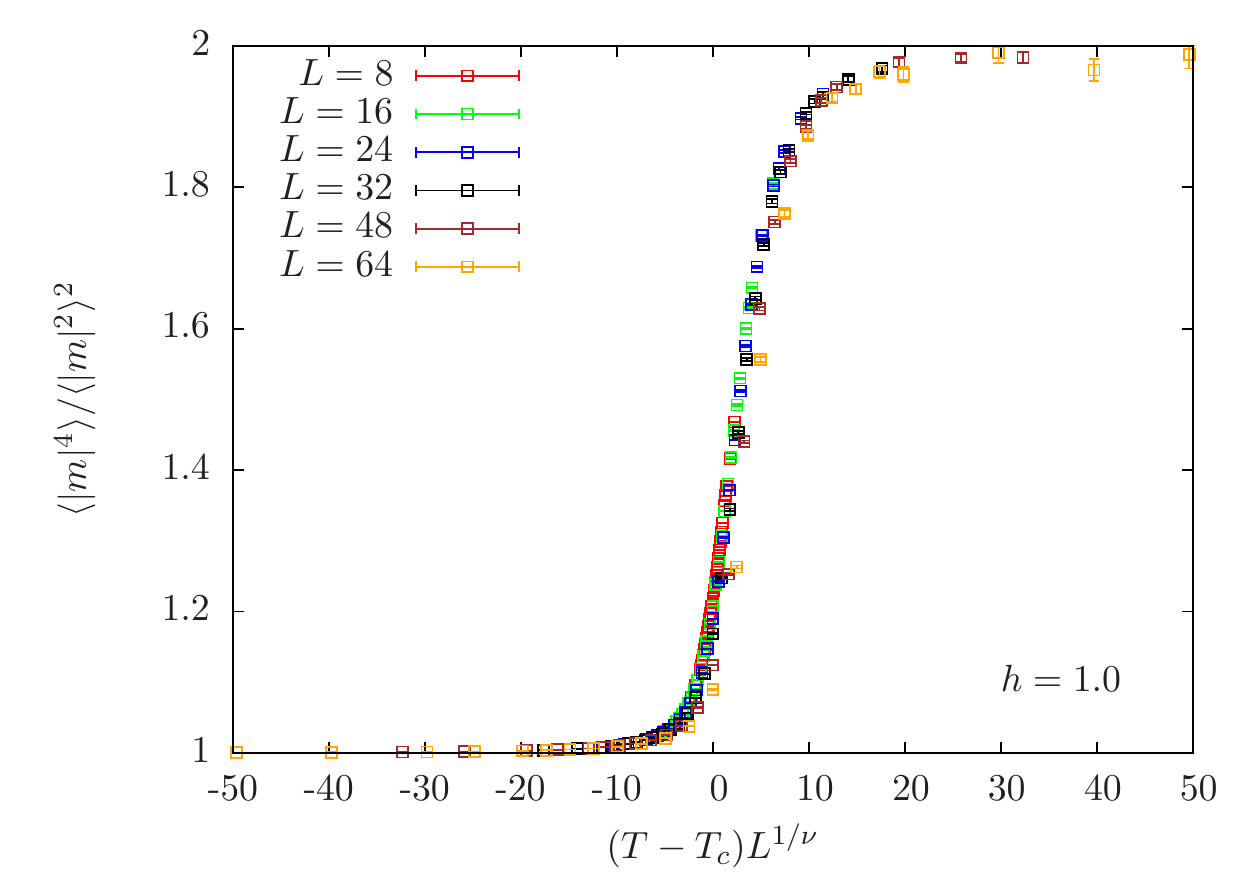}
\vspace{-2.5mm}
\caption{(Color online) 3d XY model with four-fold anisotropic field: collapse of the order parameter $\langle |m| \rangle$ (top panel, for $h=0$) and the Binder cumulant $B$ (bottom panel, for $h=1$), according to the scaling forms mentioned in the text. For estimates
on overall error-bars, refer to Table \ref{table2}.
\label{fig:3dxy_mB}
}
\end{figure}

We first locate the critical points by a standard scaling analysis for four values of the anisotropy field. To this end, we define the vector order parameter  $\vec{m} = (m_x,m_y)= \frac{1}{L^3}\sum_{\vec{r}} (\cos(\theta_{\vec{r}}) , \sin(\theta_{\vec{r}}) )$. We measure $\langle |m|\rangle$ (where $|m| \equiv \sqrt{\vec{m}^2}$) and the Binder cumulant $B = \langle (\vec{m}^2)^2\rangle/\langle \vec{m}^2\rangle^2$. We also compute the ratio $R$ of correlation functions at fixed distance $R=C_{L/2}/C_{L/4}$, where 
$C_{\ell} = \frac{1}{L^3} \sum_{\vec{r}}\langle e^{i \theta_{\vec{r}+\vec{r}_{\ell}} - i \theta_{\vec{r}}} \rangle$ and $\vec{r}_{\ell} = (\ell,\ell,\ell)$. The two dimensionless observables $B$ and $R$ are expected to satisfy the standard scaling forms $B=f_{B}((T-T_c)L^{1/\nu})$ and $R=f_{R}((T-T_c)L^{1/\nu})$ in the vicinity of a second-order critical point. Similarly, we also expect the scaling form $\langle |m| \rangle = L^{\beta/\nu} f_{m}((T-T_c)L^{1/\nu})$.

We employ this strategy at four values of the anisotropy field: $h_4=0,0.5,1.0,2.0$ and present typical results for these observables in Figs.~\ref{fig:3dxy_BR} and \ref{fig:3dxy_mB}. Fitting to the above forms allows to determine the transition
temperature $T_c(h_4)$ reasonably accurately for each of the values of $h_4$ studied. Results of our fits for $T_c(h_4)$, critical exponents and amplitudes are given in Table \ref{table2}. They clearly confirm that the universality class of the 3d XY model is unchanged by adding a $Z_4$ anisotropic field, {\it i.e.} it is an irrelevant perturbation at the critical point. Note as well how little $T_c$ changes as a function of $h_4$.

\begin{table*}
     \begin{ruledtabular}

\begin{tabular}{|c|c|c|c|c|c|c|c|c|c|c|}
\rule[-1.2ex]{0pt}{0pt}   & \multicolumn{4}{c|}{$\langle |m| \rangle$} & \multicolumn{3}{c|}{Binder ratio $B$} & \multicolumn{3}{c|}{Correlation ratio $R$}  \\
    
    \hline
$h$ & $T_c$ & $\nu$  & $\beta/\nu$ & $f_m(0)$ & $T_c$ & $\nu$  & $f_{B}(0)$ & $T_c$ & $\nu$  & $f_{R}(0)$  \\
\hline
0.0 & 2.201(1) & 0.667(2) & 0.515(1) & 1.106(6) & 2.202(1) & 0.675(10) & 1.2346(5) & 2.202(1) & 0.682(9) & 0.882(2)  \\
0.5 & 2.202(1) & 0.666(3) & 0.51(1) & 1.09(5) & 2.202(1) & 0.676(8) & 1.2365(30) & 2.203(1) & 0.671(1) & 0.882(2) \\
1.0 & 2.205(1) & 0.665(4) & 0.514(4) & 1.103(10) & 2.204(1) & 0.671(5) & 1.2377(4) & 2.205(1) & 0.67(1) & 0.883(2) \\
2.0 & 2.212(1) & 0.657(3) & 0.520(5) & 1.13(2) & 2.211(1) & 0.6572(10) & 1.2458(11) & 2.212(1) & 0.665(13) & 0.884(2)  \\
\end{tabular}
     \end{ruledtabular}
\caption{ \label{table2}
Estimates of critical temperature, exponent and amplitudes resulting from the finite-size scaling analysis of order parameter $\langle |m| \rangle$, Binder cumulants $B=\langle (\vec{m}^2)^2 \rangle / \langle \vec{m}^2 \rangle^2$ and correlation ratio $R=C_{L/2}/C_{L/4}$.
Error bars were determined from the spread on extracted fit parameters  depending
on critical window size, minimum system sizes included, 
or degree of polynomial for the universal scaling functions, with $\chi^2$ per degree of freedom always
$\lesssim 1.5$.
}
     \end{table*}

\begin{figure}[b]
\includegraphics[width=0.8 \hsize,angle=0]{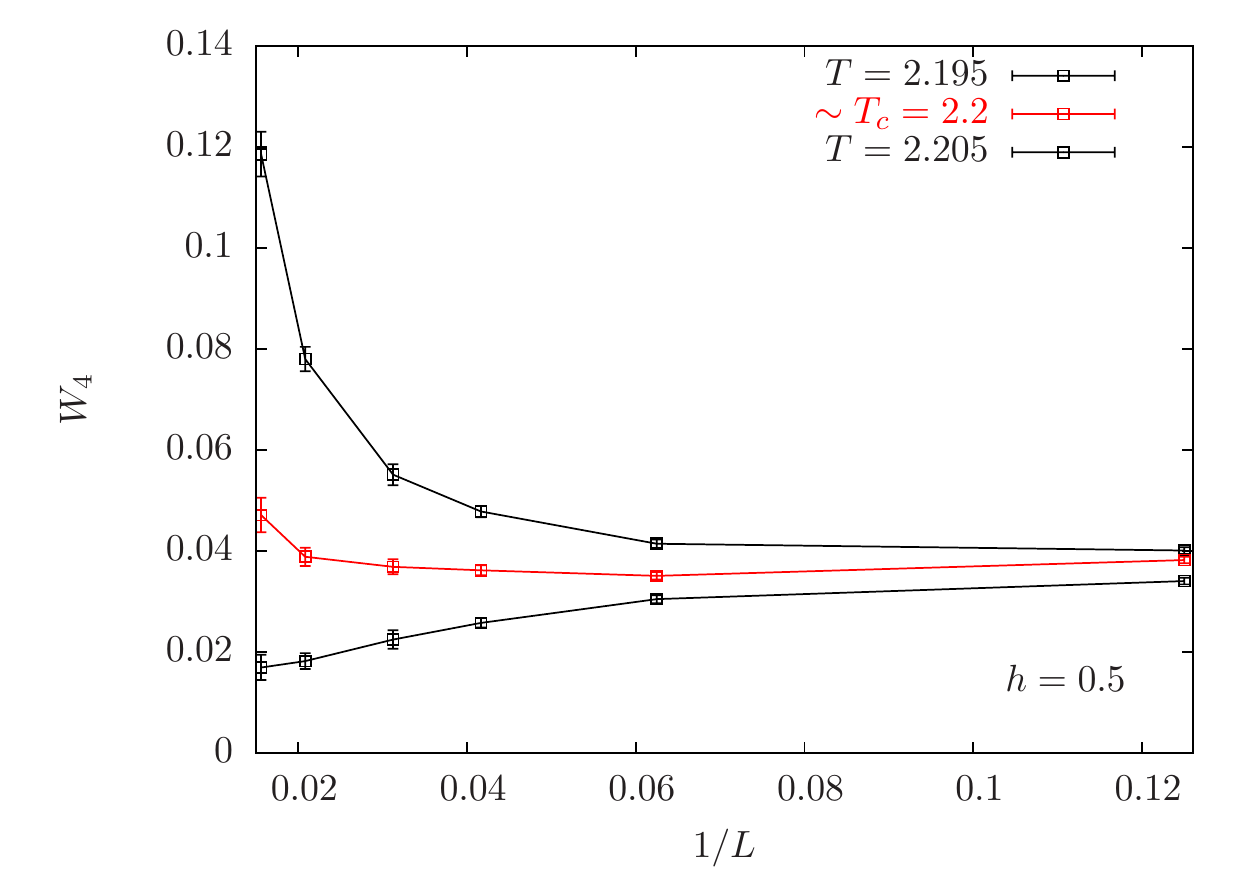}
\includegraphics[width=0.8 \hsize,angle=0]{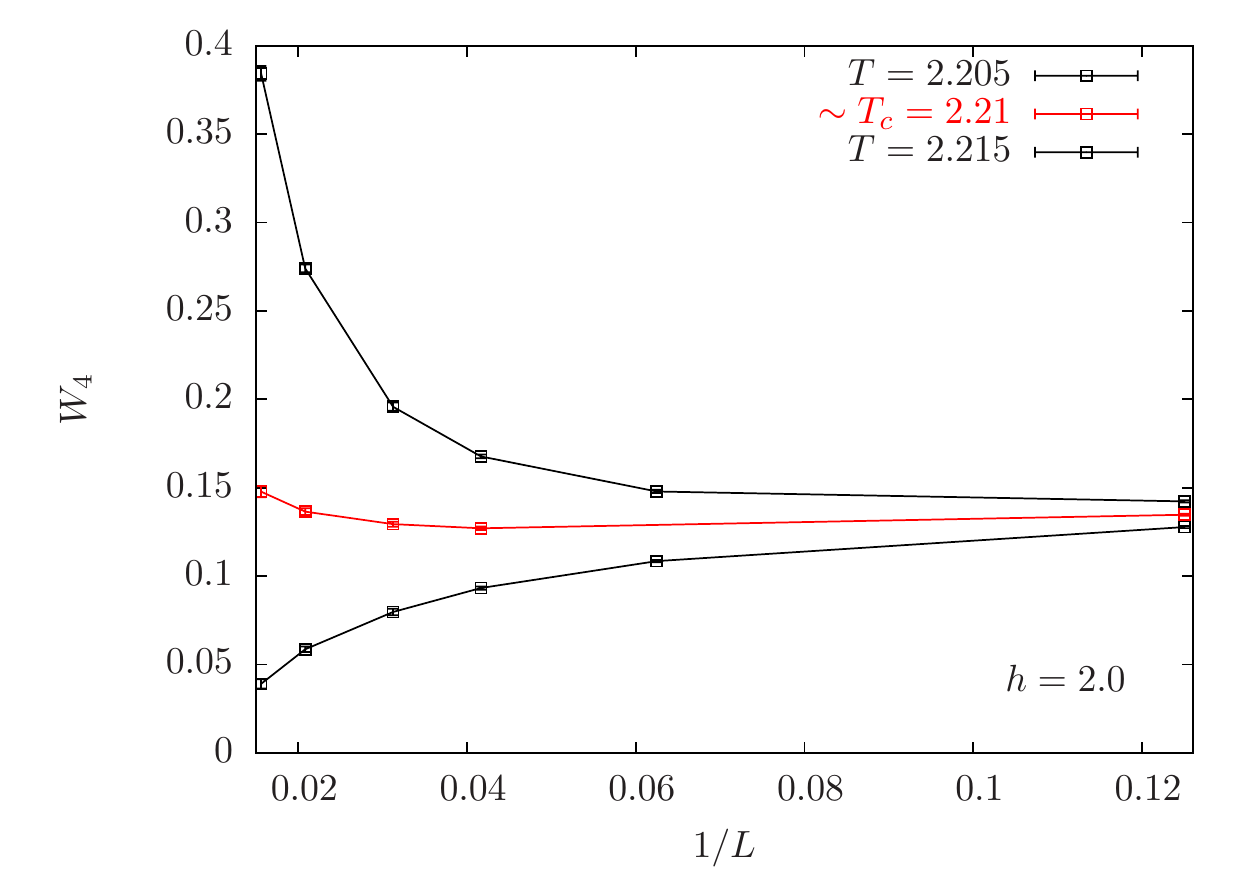}
\caption{(Color online) System-size dependence of the anisotropy parameter $W_4$ close to criticality for three different temperatures: above, below and very close to the critical temperature $T_c$. Top panel: $h=0.5$, bottom panel: $h=2$.
\label{fig:3dxy_W4}
}
\end{figure}

Armed with this knowledge, we now study $W_4$, a dimensionless
measure of $4$-fold anisotropy in the vicinity of this critical point. We define it analogously to our definition of $W_3$ for the N\'eel-VBS transition:
$W_4=\int d\vec{m} P(\vec{m}) cos(4\theta_m)$
with $P(\vec{m})$ the normalized probability distribution of the order parameter, and $\theta_m=\arctan(m_y/m_x)$ its phase, as measured during the Monte Carlo run. In Fig.~\ref{fig:3dxy_W4}, we show the size dependence of $W_4$ close to the critical point for two different values of $h_4$ (similar results are obtained for the third non-vanishing value of the field studied in our simulations). 

Whereas the anisotropy quantifier $W_4$ increases (towards its limiting value $1$) with system size below the critical temperature, it tends to vanish with system size for temperature above $T_c$. At criticality, the anisotropy $W_4$ appears to be essentially constant (and non-zero), within our range of system sizes for all nonzero $h_4$. We also find that this critical value $W_{4c}$ increases significantly with increasing $h_4$ (see Fig.~\ref{fig:3dxy_W4}). A finite-size scaling analysis of this behavior, employing some assumptions about the finite-size scaling form, is also reported in Appendix~\ref{sec:appA}, and confirms this more elementary analysis. We have also studied (see Appendix~\ref{sec:appB}) the analogous quantities for $3-$ and $5-$fold anisotropies and find that this unusual behavior is specific to the $4-$fold case.

Our results in the $Z_4$ case for this better-understood classical problem are thus entirely analogous to our results for  $W_3$ at the N\'eel-columnar VBS phase boundary. As in that case, this anisotropy coexists with other critical properties being well-fit by standard $3d-XY$ exponents. Given that $Z_4$ anisotropy is known to be weakly irrelevant at the three-dimensional $XY$ transition, this leads us to suggest that three-fold anisotropy is
also weakly-irrelevant at the N\'eel-columnar VBS transition on the honeycomb lattice.
 
\section{Outlook}
\label{sec:conc}
We close with a brief discussion of a possible avenue for further progress. 
It would be desirable to have a model system where  the bare value of the three-fold anisotropy in the phase of the VBS order
parameter $\psi$ could be tuned by hand. This would be analogous to tuning $h_4$ in the classical
three-dimensional $XY$ model. 

To achieve this, we begin with the observation that the honeycomb lattice quantum dimer model with ring-exchange on hexagonal plaquettes and {\em no inter-dimer interactions}
is known~\cite{moessner_sondhi_chandra} to order in a plaquette VBS state, corresponding to the values
$(2m+1)\pi/3$ ($m=0,1,2$) for the phase of the VBS order parameter $\psi$. The anisotropy in the phase of $\psi$ in this plaquette-ordered VBS state is thus exactly the opposite of the anisotropy in the columnar-ordered VBS phase (which corresponds to values $2\pi m /3$ for the phase of $\psi$).

Next, we note that it is possible
to write down a six-spin interaction term in a SU($N$) spin model which, for large enough $N$, mimics the ring-exchange term of the honeycomb
lattice dimer model. 
This term, given below, is the honeycomb lattice generalization
of similar constructions employed recently\cite{Kaul_2014} on the square lattice:
\begin{equation}
-R_3 \sum_{\langle i j k l m n\rangle} (| (ij) (kl) (mn) \rangle \langle (jk) (lm) (ni)| + h.c.).
\end{equation}
Here, the sum is over all such plaquettes of the honeycomb lattice labelled by $\langle i j k l m n \rangle$ with vertices labeled cyclically, and $| (ij) (kl) (mn) \rangle$ is the state in which (SU($N$)) spins $i$ and $j$ form a (SU($N$)) singlet (similarly for spins $k$ and $l$, and $m$ and $n$). In the large-$N$ limit, this reduces
to a ring-exchange term on each plaquette.

With this motivation, we expect that a non-zero $R_3$ will counter the columnar phase anisotropy seen at the critical point of the SU($2$) invariant $J-Q_3$ model and allow us
to tune the value of $W_3$ while leaving other critical properties unchanged.
Thus, we conjecture that the SU($2$) invariant $J-Q_3-R_3$ model (employing the
$R_3$ term defined above) provides a promising setting in which one can tune the bare value of the anisotropy in the phase of $\psi$, and explicitly check the idea that this three-fold anisotropy
is a weakly irrelevant variable at the N\'eel-columnar VBS transition.
In addition, it may even be possible to change the character of the ordered state (from
columnar to plaquette VBS) if $R_3$ dominates over $Q_3$. It should
be possible to confirm these ideas using projector QMC simulations
of this $J-Q_3-R_3$ model, and we hope to return to this in future work.

\vspace{-3mm}
\acknowledgements{This work was made possible by research
support from the Indo-French Centre for the Promotion of Advanced
Research (IFCPAR/CEFIPRA) under Project 4504-1 and DST grant DST-SR/S2/RJN-25/2006, and performed using computational resources from GENCI (grant x2014050225), CALMIP (grant 2014-P0677) and of the Dept. of Theoretical Physics of the TIFR. SP is grateful to the Dept. of Theoretical Physics of the TIFR for hospitality during part of
this work. In the final stages of this work, SP was also supported by NSF grant
DMR-1056536.}
\vspace{-2mm}

\appendix

\section{Finite-size scaling analysis of the dimensionless anisotropy quantifier}
\label{sec:appA}

To supplement the $W_3$ versus $L$ behavior at fixed $Q_2$ that we looked at in the main text, we perform a finite-size
scaling analysis based on the scaling theory of Lou \emph{et al} \cite{Lou_Sandvik_Balents}.
Ref.~\onlinecite{Lou_Sandvik_Balents} studied the classical 3d $XY$ model in presence of a $Z_q$ anisotropy field, 
which is a dangerously irrelevant operator at criticality for $q \geq 4$, and proposed a scaling form for the dimensionful anisotropy order parameter as $\langle m_q \rangle = L^{-\beta/\nu} f_{m_q} ((T-T_c)L^{1/\nu_q})$, an extension of the $XY$ order
parameter scaling form $\langle m \rangle = L^{-\beta/\nu} f_m((T-T_c)L^{1/\nu})$. $\nu_q$ is the exponent associated with a length scale below which the order parameter distribution appears isotropic, even below $T_c$. We have $\nu_q > \nu$, as this length scale diverges faster than the ferromagnetic correlation length (see the analogy with the VBS anisotropy length scale in the theory of deconfined criticality~\cite{Senthil_etal_PRB,Senthil_etal_Science}). Ref.~\onlinecite{Lou_Sandvik_Balents} related $\nu_q/\nu$ to the scaling dimension of the anisotropy field, but we note that in a recent work this relation was questioned~\cite{okubo}.

\begin{figure}[t]
\includegraphics[width=0.8\columnwidth,angle=0]{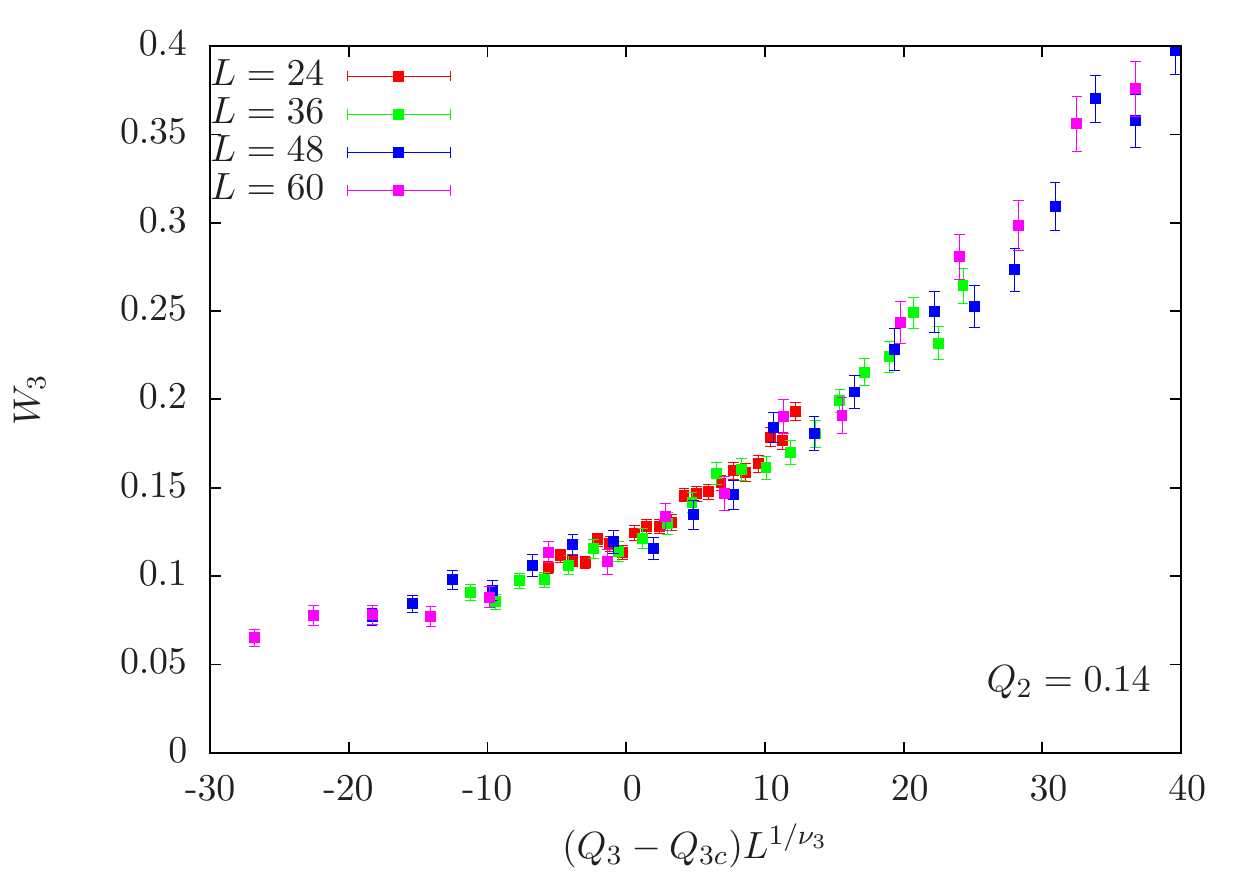}
\includegraphics[width=0.8\columnwidth,angle=0]{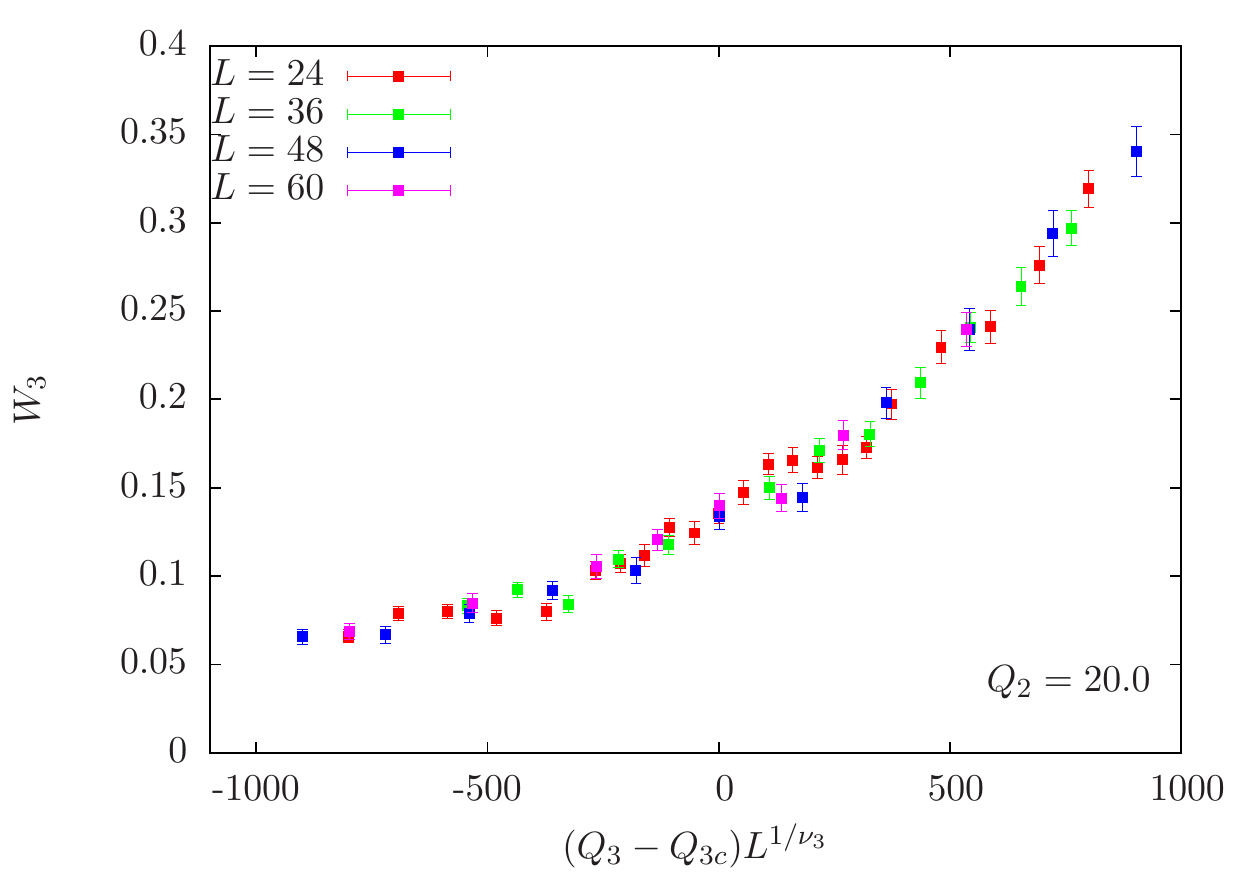}
\vspace{-2mm}
\caption{(Color online) Scaling collapse plots according to the scaling form $W_3=g_{W_3}((Q_3-Q_{3c})L^{1/\nu_3})$ for the
dimensionless anisotropy quantifier $W_3$ in the $J-Q_2-Q_3$ model
for $Q_2=0.14$ (top panel) and $Q_2=20$ (bottom panel). 
For the fits, similar to Sec. \ref{subsec:exponents_and_scaling_forms}, 
a particular choice of critical window, minimum system size included, and 
order of universal function was taken here which gave $\chi^2$ per degree of freedom
equal to 1.26
and 1.31 for the plots respectively.
\label{fig:W3_collapse}
}
\end{figure}

{\it $J-Q_2-Q_3$ model --- } In our case of the dimensionless anisotropy order parameter $W_3$, we can assume following Ref.~\onlinecite{Lou_Sandvik_Balents} a similar scaling form $g_{W_3}((Q_3-Q_{3c})L^{1/\nu_3})$ for fixed $Q_2$, without further assumption on $\nu_3$. Fig. \ref{fig:W3_collapse} shows examples of this scaling analysis and Tab. \ref{table:W3_fits_jq3q2}
summarizes the results of the corresponding fits. 

\begin{table}
\begin{tabular}{|c|c|c|c|c|}
\hline
$Q_2$ & $Q_{3c}$ & $\nu_3$  & $g_{W_3}(0)$ \\
\hline
0.0 & 1.183(2) & 0.57(2) & 0.115(6) \\
0.14 & 1.485(1) & 0.58(1) & 0.120(3) \\
0.60 & 2.485(1) & 0.56(2) & 0.129(2) \\
0.85 & 3.027(2) & 0.56(2) & 0.128(3)\\
20.0 & 45.00(3) & 0.57(1) & 0.134(2) \\
\hline
\end{tabular}
\caption{ \label{table:W3_fits_jq3q2}
Results of finite size scaling analysis for the
dimensionless anisotropy quantifier $W_3$ for the $J-Q_2-Q_3$ model.
Error bars were determined again using the same protocol as in 
Sec. \ref{subsec:exponents_and_scaling_forms} of the main text 
(see Table \ref{table1}).
}
\end{table}

We see that the critical point $Q_{3c}$
extracted from the scaling analysis is again in agreement with those
gotten from other analyses (Sec. \ref{subsec:exponents_and_scaling_forms}).
We again find the same conclusions as that from visual inspection of
$W_3$ versus $L$ behavior: there is a finite value of $W_{3c}=g_{W_3}(0)$ at the 
critical point for all $Q_2$, which furthermore seems to slightly increase  with $Q_2$. 
Finally, within our precision, it is not possible to positively confirm that the extracted value of $\nu_3$ is larger than $\nu$ (the two exponents are essentially equal within error bars): independent of the exact relation between the two~\cite{Lou_Sandvik_Balents,okubo}, this indicates that $3-$fold anisotropy is only very slightly irrelevant, consistent with a non-vanishing $W_{3c}$ within our system size range.

\begin{figure}[t]
\includegraphics[width=0.8\columnwidth,angle=0]{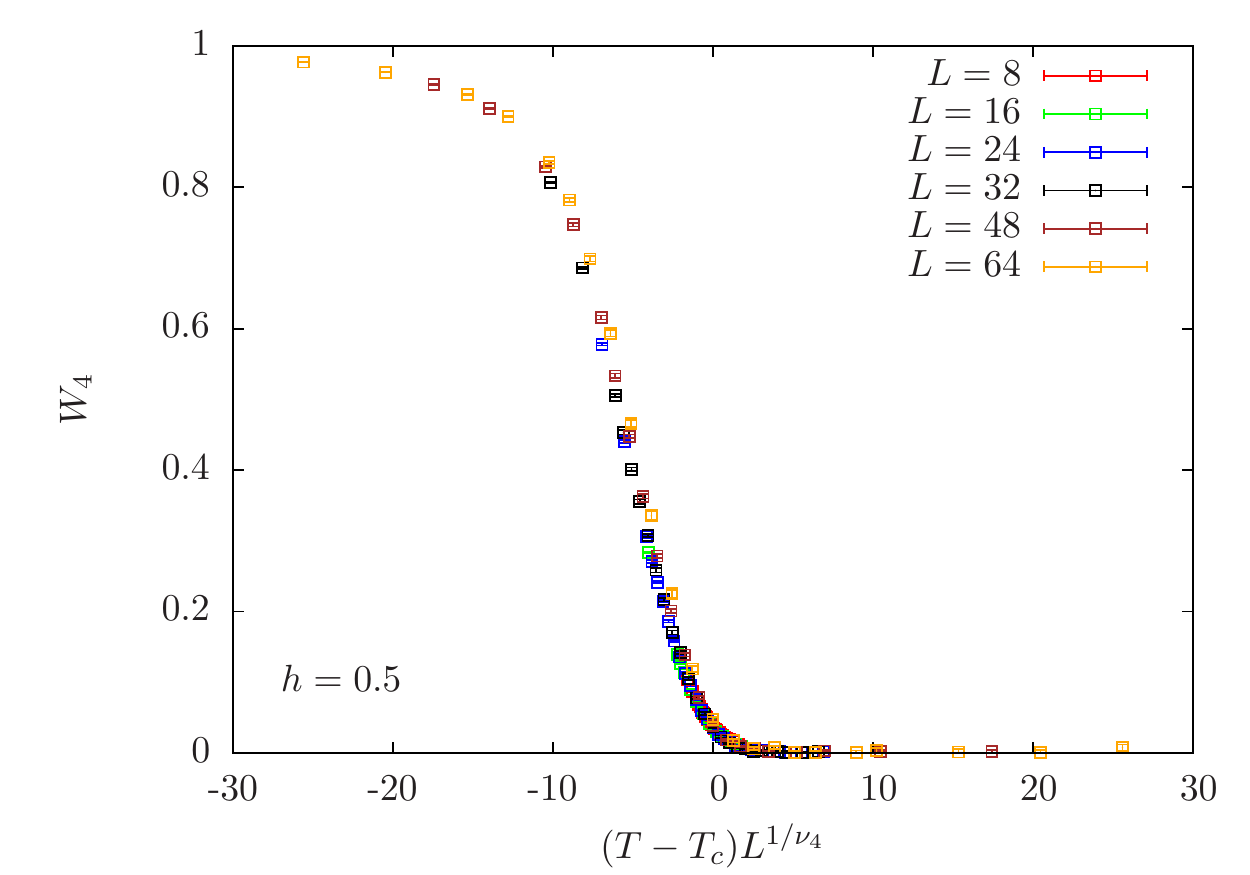} 
\includegraphics[width=0.8\columnwidth,angle=0]{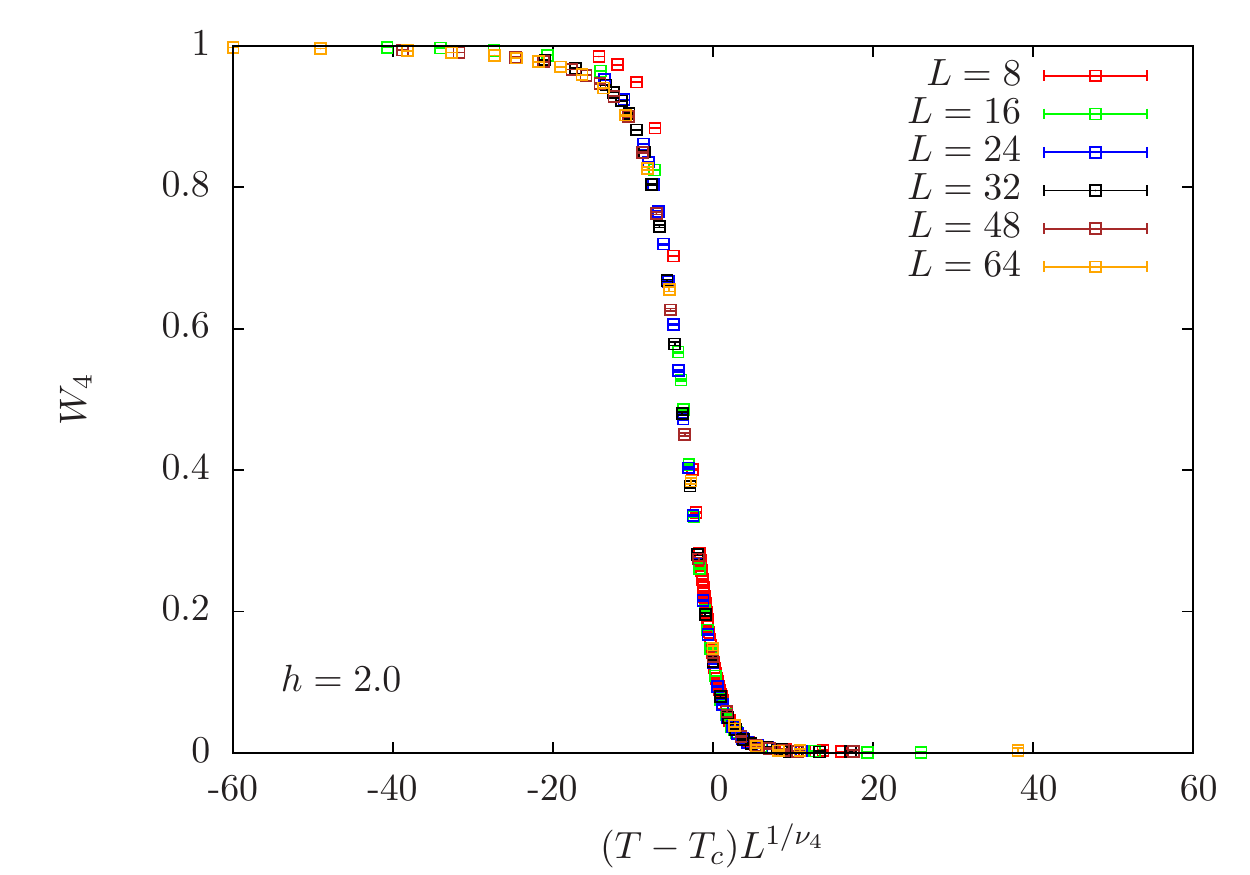}
\caption{(Color online) Scaling collapse plots according to the scaling form $W_4=g_{W_4}((T - T_c)L^{1/\nu_4})$ for the
dimensionless anisotropy quantifier $W_4$ in the 3d XY model with $4-$fold anisotropy field for $h=0.5$ (top panel) and $h=2$ (bottom panel). 
For estimates
on overall error-bars, refer to Table \ref{table:W4_fits_3dxy}.}
\label{fig:W4_3dxyq4}
\end{figure}

{\it 3d XY model with 4-fold anisotropy field --- }  We perform the same analysis for the anisotropy quantifier $W_4$ of the 3d XY model. In Fig. \ref{fig:W4_3dxyq4}, we show the scaling collapse for $W_4$ with the scaling form $W_4=g_{W_4}((T - T_c)L^{1/{\nu_4}})$ as the anisotropy field $h_4$ is varied.
Table \ref{table:W4_fits_3dxy} summarizes the results of the scaling analyses. 

\begin{table}[h]
\begin{tabular}{|c|c|c|c|}
\hline
$h$ & $T_c$ & $\nu_4$  & $f_{W_4}(0)$\\
\hline
0.5 & 2.202(2) & 0.76(10) & 0.031(4) \\
1.0 & 2.204(1) & 0.70(2) & 0.062(4) \\
2.0 & 2.211(1) & 0.665(20) & 0.120(1) \\
\hline
\end{tabular}
\caption{ \label{table:W4_fits_3dxy}
Results of finite size scaling analysis for anisotropy quantifier $W_4$ for the 3d XY model with $4-$fold anisotropic field. Error bars were determined with the same procedure as in the main text (see Table \ref{table1}).}
\end{table}

We find again the critical temperature $T_c$
is in agreement with those extracted from other order parameters (Sec. \ref{sec:3dxy}) and changes very little with $h_4$, as already mentioned. This analysis confirms that $W_4$ takes a clearly non-zero value $W_{4c}=f_{W_4}(0)$ at the critical point, which logically increases with $h_4$. In this case, we are able to confirm that $\nu_4 > \nu$ as found in Ref. \onlinecite{Lou_Sandvik_Balents} except for the largest field $h=2$ where this relation is only marginally verified (this can be expected as we probably need larger systems when anisotropy is stronger).

\section{$3d$ $XY$ model with $3-$ and $5-$fold anisotropic fields }
\label{sec:appB}

Here we show that a nearly-constant critical anisotropy is specific to the 3d XY model with $4-$fold anisotropic field by studying the same model with a $3-$ and $5-$fold anisotropy field, replacing the term  $- h_4 \sum_{\vec{r}}\cos(4 \theta_{\vec{r}})$ by  $- h_q \sum_{\vec{r}}\cos(q \theta_{\vec{r}})$ with $q=3,5$ in Eq.~\ref{eq:3dxy_hamiltonian}. 
We again compute the Binder cumulant and the anisotropy quantifiers $W_3$ and $W_5$ adapting the above definitions.

\begin{figure}[b]
\includegraphics[width=0.8\columnwidth,angle=0]{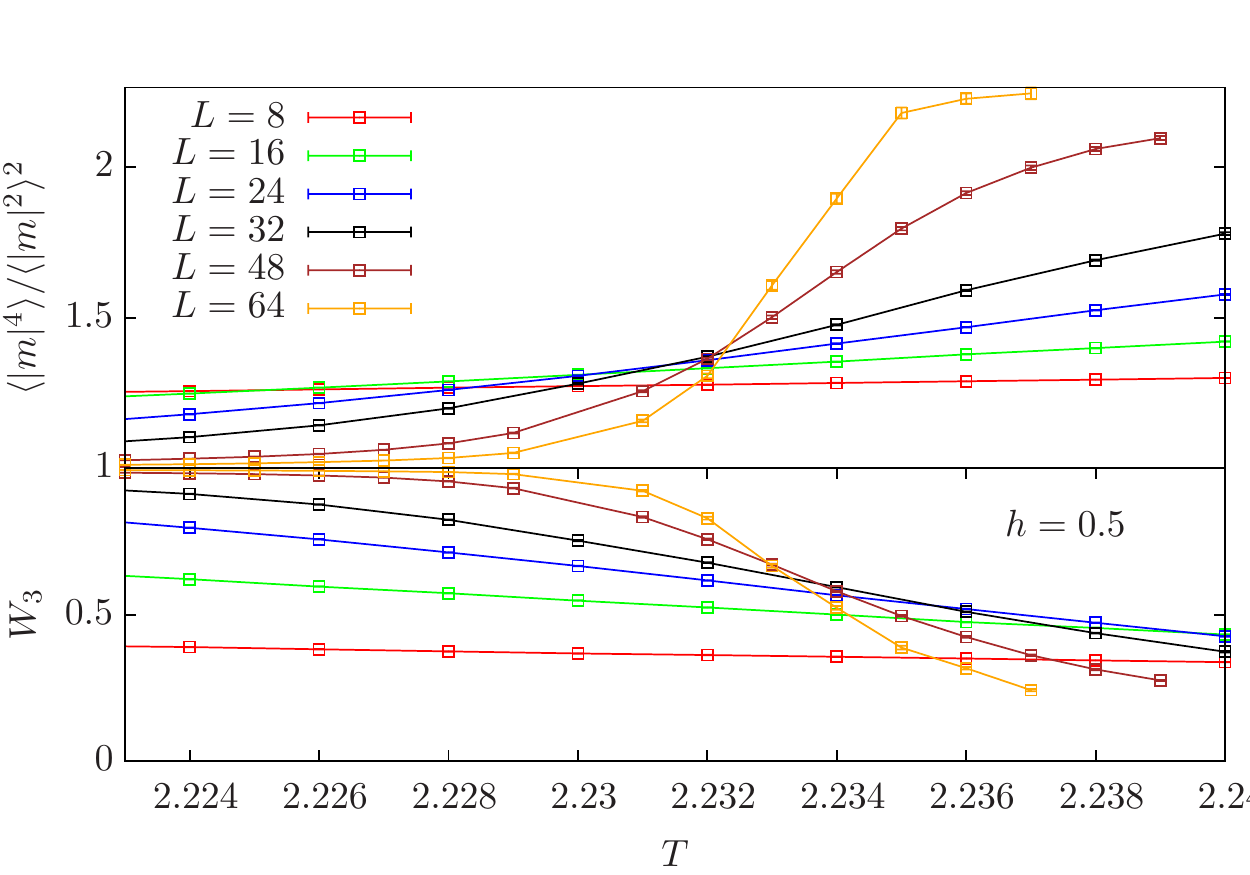}
\includegraphics[width=0.8\columnwidth,angle=0]{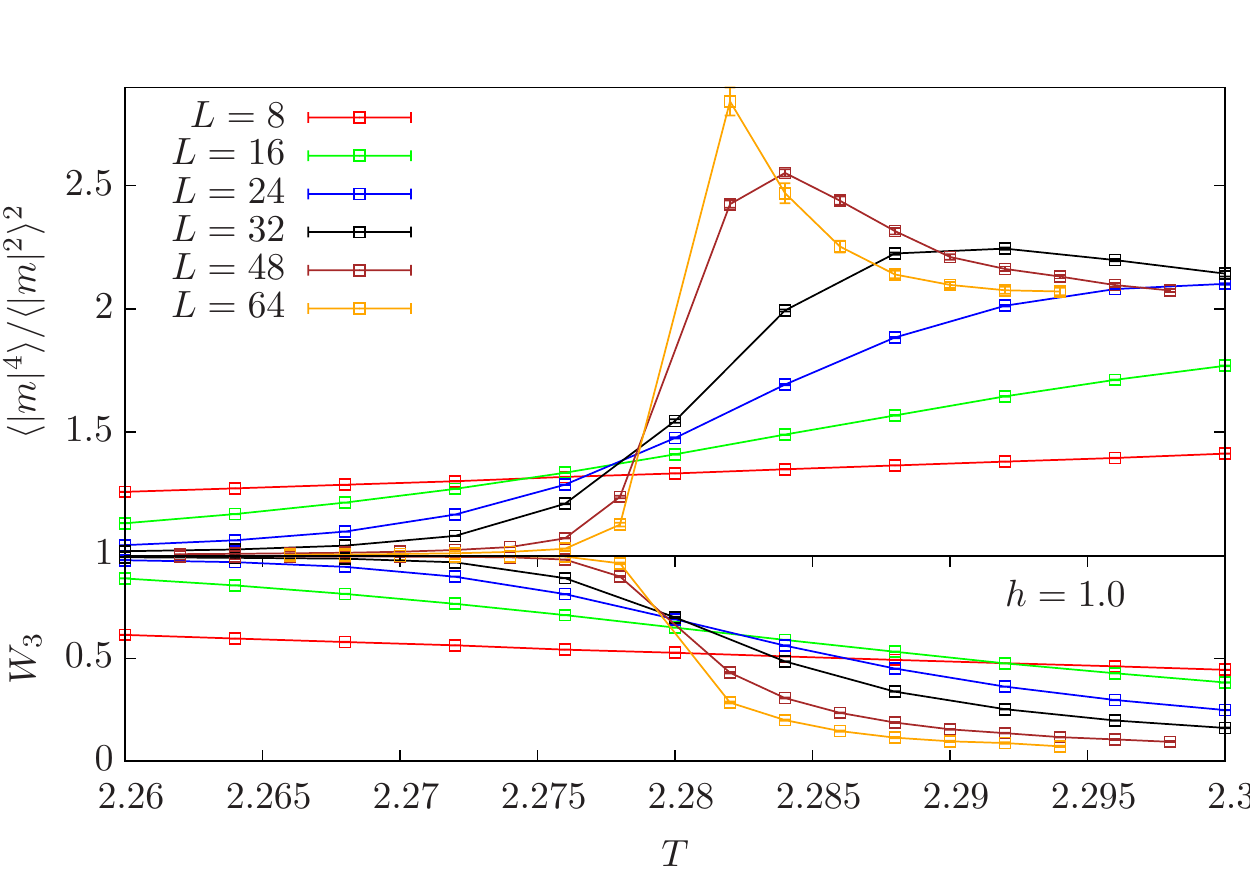}
\vspace{-2mm}
\caption{(Color online) 3d XY model with $3-$fold anisotropic field: temperature dependence of the Binder cumulant (top panels) and $3-$fold anisotropy quantifier $W_3$ (bottom panels) for two different values of $h_3={0.5,1.0}$.}
\label{fig:binder_W3_3dxyq3}
\end{figure}

{\it $q=3$ case --- } We know that the anisotropy is relevant here, rendering the transition first-order. This is clearly seen in the top panels of Fig. \ref{fig:binder_W3_3dxyq3} where, for two different field values, the Binder cumulant show significant drifts in the crossing point between two consecutive sizes. The bottom panels show the temperature dependence of $W_3$, which also show drifting pseudo-crossing points. The clear increase with $L$ of $W_3$ nearest to the transition temperature
where the pseudo-crossing in the Binder cumulant is located indicates that anisotropy is relevant at criticality. Note as well how the value of $T_c$ is substantially modified by $h_3$.

\begin{figure}
\includegraphics[width=0.8\columnwidth,angle=0]{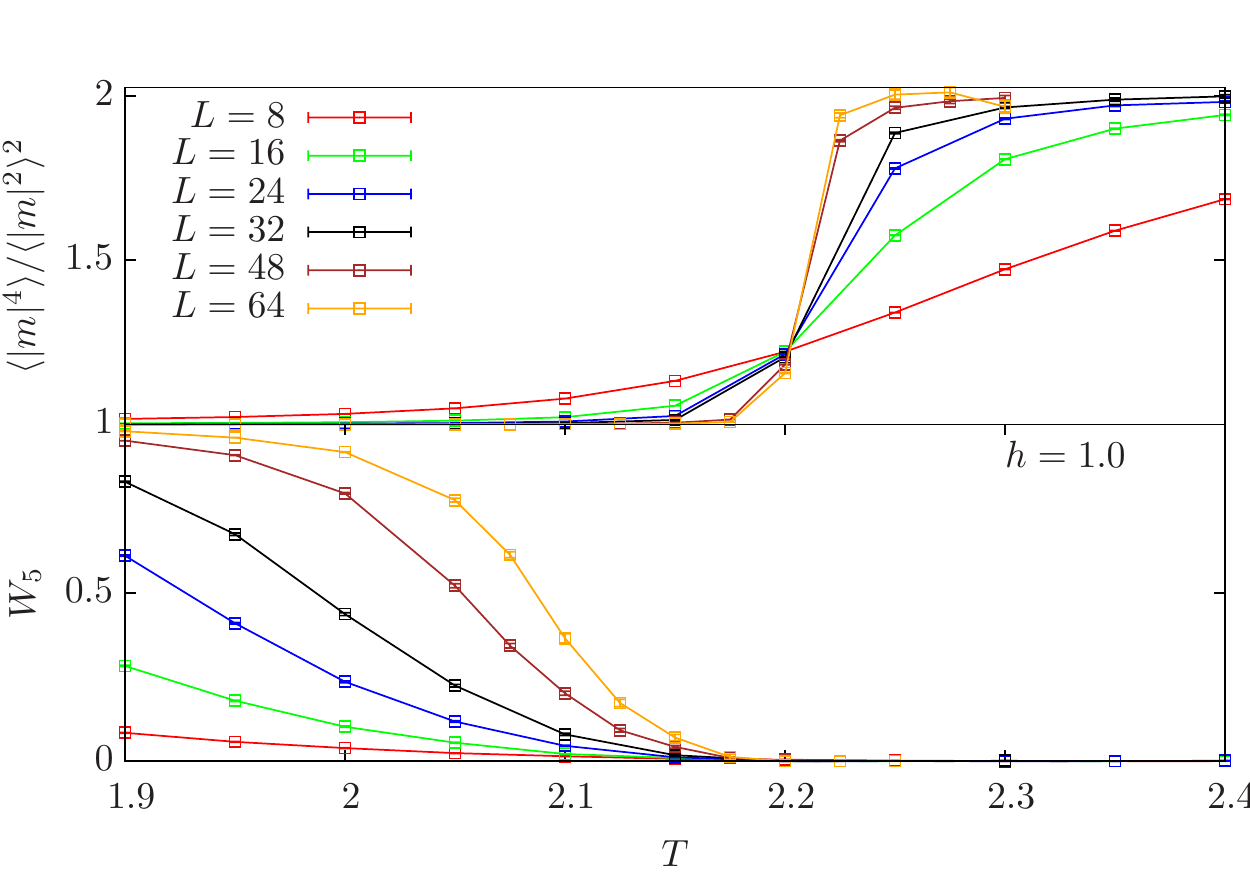}
\includegraphics[width=0.8\columnwidth,angle=0]{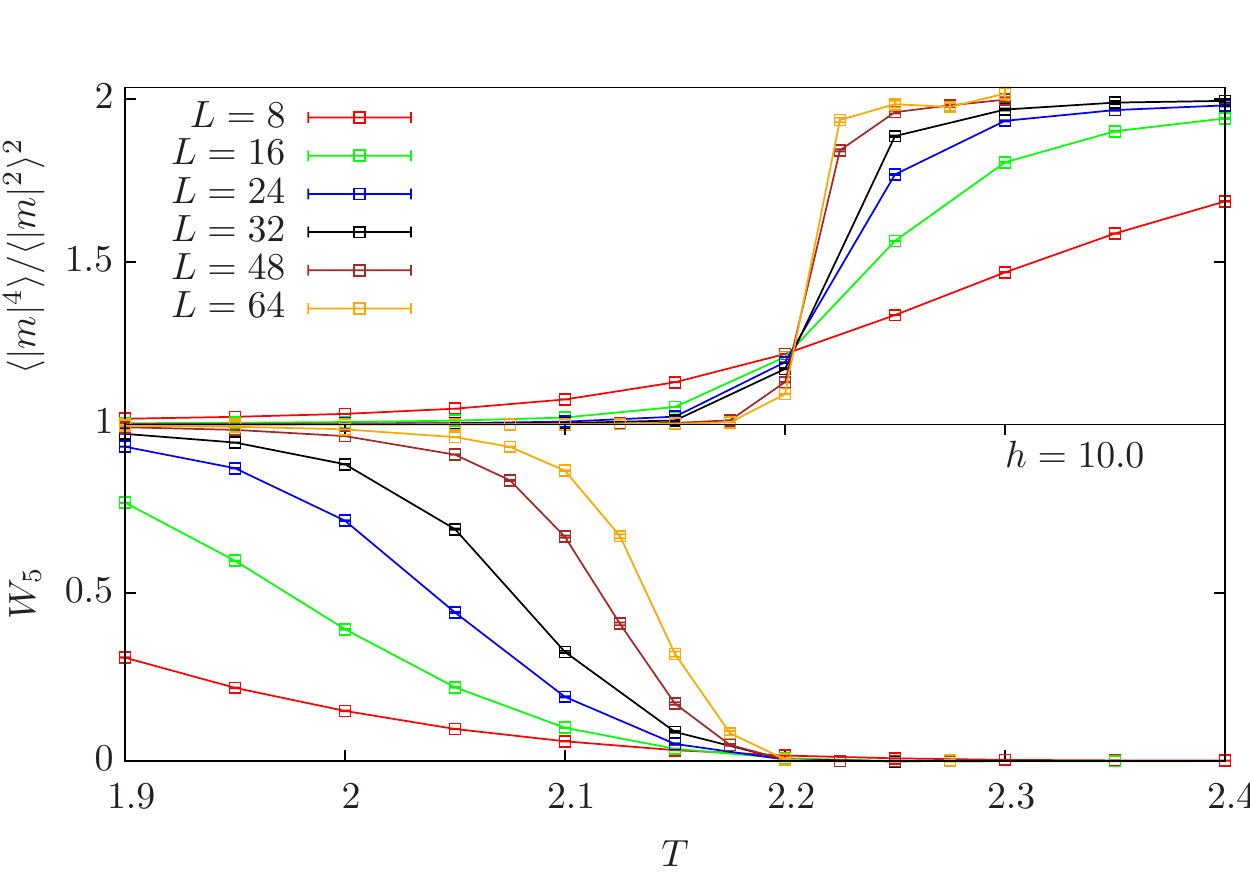}
\caption{Color online) 3d XY model with $5-$fold anisotropic field: temperature dependence of the Binder cumulant (top panels) and $5-$fold anisotropy quantifier $W_5$ (bottom panels) for two different values of $h_5={1,10}$.}
\label{fig:binder_W5_3dxyq5}
\end{figure}

{\it $q=5$ case --- } Anisotropy is irrelevant here and the second order nature of the transition is revealed by
the nice monotonic crossing behavior of the Binder cumulant in the top panels of Fig.~\ref{fig:binder_W5_3dxyq5} for two different values of $h_5$. There is no observable drift in $T_c$ even when $h_5$ changes by a factor of $10$ -- in fact, one observes that the Binder cumulant are essentially the same, indicating the strong
irrelevancy of $5-$fold anisotropy. The bottom panels of Fig.~\ref{fig:binder_W5_3dxyq5} show the size and temperature dependence of $W_5$, which as expected clearly goes to zero at the critical point. We performed a finite-size scaling analys of the data (not shown) which yield the expected results, such as non-drifting $T_c$, $\nu_5 > \nu $, $f_{W_5}(0)=0$ and the correct 3d XY value for $\nu$.

\end{document}